\RequirePackage{fix-cm}
\documentclass[smallextended]{svjour3}       
\smartqed  
\usepackage{algorithm}
\usepackage{algorithmicx}
\usepackage{algpseudocode}
\usepackage{graphicx}
\usepackage{subfigure}
\usepackage{float}
\usepackage{color}
\usepackage[tbtags]{amsmath}
\usepackage{amsfonts}
\usepackage{epsfig}
\usepackage{graphicx}
\usepackage{arydshln}
\usepackage{verbatim}
\usepackage{subfigure}
\usepackage{enumerate}
\usepackage{rotating}
\usepackage{threeparttable}
\usepackage{caption}
\usepackage{epsfig}
\usepackage{cite}
\journalname{Encyclopedia of Electrical and Electronics Engineering}
\date{}
\newtheorem{mydef}{Definition}
\newtheorem{myprop}{Proposition}

\newtheorem{myexm}{Example}

\begin{document}

\title{Estimation and Verification of Partially-Observed Discrete-Event Systems}

\author{Xiang Yin}

\institute{X. Yin \at
 Department of Automation, Shanghai Jiao Tong University, Shanghai 200240, China.\\
              \email{yinxiang@sjtu.edu.cn}           
}

\maketitle

\section{Introduction}
State estimation is one of the most fundamental problems in the systems and control theory.
In many real-world systems, it is not always possible to  access the full information of the system due to  measurement noises/uncertainties and the existence of adversaries.
Therefore, how to \emph{estimate} the state of the system is crucial when one wants to make decisions based on the  limited and incomplete information.

Given a system, one of the most important questions is to prove/disprove the correctness of the system with respect to  some desired requirement (or specification).
In particular, we would like to check the correctness of the system in a formal manner in the sense that the checking procedures are algorithmic and  results have provable correctness guarantees. Such a formal satisfaction checking problem is referred to as the \emph{verification problem}.
Performing formal  property verification is very important for many complicated but safety-critical infrastructures.

This article considers  state estimation and verification problems for an important class of man-made cyber-physical systems called Discrete-Event Systems (DES).
Roughly speaking, DES are dynamic systems with discrete state-spaces and event-triggered dynamics.
DES models are widely used in the study of complex automated systems where the behavior is inherently event-driven,
as well as in the study of discrete abstractions of continuous, hybrid, and/or cyber-physical systems.
Over  the past decades, the theory of DES has been successfully applied to many real-world problems, 
e.g., the control of automated systems, fault diagnosis/prognosis and information-flow security analysis.

The main purpose of this article is to provide a tutorial and an overview of state estimation techniques and their related property verification problems for \emph{partially-observed DES}.
Specifically, we focus on  the verification of \emph{observational properties}, i.e.,  information-flow properties whose satisfactions are based on the observation of the system.
The outline of this article is as follows: 
\begin{itemize}
  \item
  In Section~\ref{sec:2}, we briefly introduce some necessary terminologies and the partially-observed DES model considered in this article.
  \item
  In Section~\ref{sec:3}, we introduce three types of state estimation problems, namely, current state estimation, initial state estimation and delayed state estimation.
  Then we provide state estimation techniques  for different state estimation problems.
  \item
  In Section~\ref{sec:4}, we introduce several important observational properties that arise in the analysis of partially-observed DES.
  In particular, we consider detectability, diagnosability, prognosability, distinguishability and opacity, which cover most of the important properties in partially-observed DES.
  One feature of this section is that we study all these properties  in a uniform manner by defining them in terms of state estimates introduced in Section~\ref{sec:3}.
  \item
  In Section~\ref{sec:5}, we provide verification procedures for all properties introduced in Section~\ref{sec:4}.
  All verification procedures are summarized as detailed algorithms using  state estimation techniques provided in Section~\ref{sec:3}.
  \item
  In Section~\ref{sec:6}, some related problems and further readings on estimation and verification of partially-observed DES are provided.
\end{itemize}
The entire article is self-contained. Related references are provided in each section.

\section{Partially-Observed Discrete-Event Systems}\label{sec:2}
This section provides the basic   model of partially-observed DES and some related terminologies.
The reader is referred to the textbook \cite{Lbook} for more details on DES.

Let $\Sigma$ be a finite set of events.
A string $s=\sigma_1\dots\sigma_n,\sigma_i\in \Sigma$ is a finite sequence of events.
We denote by $\Sigma^*$ the set of all strings over $\Sigma$ including the empty string $\epsilon$.
For any string $s\in\Sigma^*$, we denote by $|s|$ its length with $|\epsilon|=0$.
A language $L\subseteq \Sigma^*$ is a set of strings.
For any string $s\in L$ in language $L$, we denote by $L/s$ the post-language of $s$ in $L$, i.e.,
$L/s:=\{w\in \Sigma^*: sw\in L\}$.
The prefix-closure of language $L$ is defined by $\overline{L}=\{ s\in \Sigma^*:\exists {w}\in \Sigma^* \text{ s.t. }sw\in L \}$;
$L$ is said to be prefix-closed if $L=\overline{L}$.

A DES is modeled as a non-deterministic finite-state automaton (NFA)
\begin{equation}
G=(X, \Sigma,\delta,X_0), 
\end{equation}
where $X$ is a finite set of states, $\Sigma$ is a finite set of events,
$\delta: X\times\Sigma \to 2^X$ is the (partial) non-deterministic transition function,
where for any $x,x'\in X,\sigma\in\Sigma$, $x'\in \delta(x,\sigma)$ means that there exists a transition from state $x$ to state $x'$ with event label $\sigma$,
and $X_0\subseteq X$ is the set of initial states.
The transition function is also extended to $\delta:X\times \Sigma^*\to 2^X$ recursively by:
(i) $\delta(x,\epsilon)=\{x\}$; and
(ii) for any $x\in X,s\in\Sigma^*,\sigma\in \Sigma$, we have  $\delta(x,s\sigma)=   \cup_{x'\in \delta(x,s)}   \delta(x',\sigma)$.
For the sake of simplicity, we write $\delta(s)=\cup_{x_0\in X_0}\delta(x_0,s)$.
We define $\mathcal{L}(G,x)=\{s\in \Sigma^*:\delta(x,s)!\}$ as the set of strings that can be generated by $G$ from state $x\in X$, where ``$!$" means ``is defined".
We define $\mathcal{L}(G):=\cup_{x_0\in X_0}\mathcal{L}(G,x_0)$ the language generated by system $G$.
System $G$ is a deterministic finite-state automaton (DFA)
 if (i) $|X_0|=1$; and (ii) $\forall x\in X,\sigma\in \Sigma:|\delta(x,\sigma)|=1$.
We say that a sequence of state $x_0x_1\dots x_n$ forms a \emph{cycle} in $G$ if (i) $\forall i=0,\dots,n-1,\exists \sigma\in \Sigma: x_{i+1}\in \delta(x_i,\sigma)$;
and (ii) $x_0=x_n$.

Let $G_1=(X_1,\Sigma_1,\delta_1,X_{0,1})$ and $G_2=(X_2,\Sigma_2,\delta_2,X_{0,2})$ be two automata.
The \emph{product} of $G_1$ and $G_2$, denoted by $G_1\times G_2$, is defined as the accessible part of a new NFA
\begin{equation}
G_1\times G_2=(X_1\times X_2, \Sigma_1\cup \Sigma_2, \delta_{1,2}, X_{0,1}\times X_{0,2}),
\end{equation}
where the transition function is defined by: for any $(x_1,x_2)\in X_1\times X_2$, for any $\sigma\in \Sigma_1\cap  \Sigma_2$, we have
\begin{equation}
\delta_{1,2}((x_1,x_2),\sigma)\!=\!
\left\{
\begin{array}{l l}
\delta_1(x_1,\sigma) \times \delta_2(x_2,\sigma) &\text{if }  \sigma\in \Sigma_1\cap\Sigma_2  \\
\text{undefined}                                             &\text{otherwise}
\end{array}
\right.
\end{equation}

In the partial observation setting, not all events generated by the system can be observed.
Formally, we assume that the event set is partitioned as follows
\begin{equation}
\Sigma=\Sigma_o\dot{\cup}\Sigma_{uo},
\end{equation}
where $\Sigma_o$ is the set of observable events and $\Sigma_{uo}$ is the set of unobservable events.
The natural projection from $\Sigma$ to $\Sigma_o$ is a mapping $P:\Sigma^*\to \Sigma_o^*$ defined recursively as follows:
\begin{align}
  P(\epsilon)=\epsilon \text{ \ and \ }
  P(s\sigma)=
  \left\{
      \begin{array}{l l}
       P(s)\sigma\quad
       &\text{if } \sigma \in \Sigma_o   \\
       P(s)\quad
       &\text{if } \sigma \notin \Sigma_o
       \end{array}
  \right.
\end{align}
That is, for any string $s\in\Sigma^*$, $P(s)$ is obtained by erasing all unobservable events in it.
We denote by $P^{-1}:\Sigma_o^*\to 2^{\Sigma^*}$ the inverse projection, i.e.,
for any $\alpha\in \Sigma^*_o$, we have $P^{-1}(\alpha)=\{    s\in \Sigma^*: P(s)=\alpha     \}$.
Note that the codomain of $P^{-1}$ is $ 2^{\Sigma^*}$ as the inverse projection of a string is not unique in general.
We also extend the natural projection to $P:2^{\Sigma^*}\to 2^{\Sigma_o^*}$ by:
for any $L\subseteq \Sigma^*$, $P(L)=\{P(s)\in \Sigma_o^*: s\in L\}$.
The inverse mapping is also extended to $P^{-1}:2^{\Sigma_o^*}\to 2^{\Sigma^*}$ analogously.

Hereafter, we make the following standard assumptions in the analysis of partially-observed DES:
\begin{enumerate}[{A}1]
  \item 
  System $G$ is live, i.e., $\forall x\in X,\exists \sigma\in \Sigma: \delta(x,\sigma)!$; and
  \item 
  System $G$ does not contain an unobservable cycle, i.e., $\forall x,x'\in X, s\in \Sigma_{uo}^*\setminus\{\epsilon\}:x'\notin \delta(x,s)$.
\end{enumerate}

\section{State Estimation under Partial Observation}\label{sec:3}

\subsection{State Estimation Problems}
Since the system is partially-observed, one important question is what do we know about the system's state when a (projected) string is observed.
This is referred to as the \emph{state estimation problem} in the system's theory.

One  of the most fundamental  state estimation problems is  the  current state estimation problem, 
 i.e., we want to estimate all possible states the system can be in currently based on the observation.

\begin{mydef}(Current-State-Estimate)
Let $G$ be a DES with observable events $\Sigma_o\subseteq \Sigma$ and $\alpha\in P(\mathcal{L}(G))$ be an observed string.
The current-state-estimate upon the occurrence of $\alpha$, denoted by $\hat{X}_G(\alpha)$, is the set of states the system could be in currently based on observation $\alpha$, i.e.,
\begin{equation}
\hat{X}_G(\alpha)=\{  x\in X: \exists x_0\in X_0,s\in \mathcal{L}(G,x_0)\text{ s.t. }x\in \delta(x_0,s)\wedge P(s)=\alpha     \}.
\end{equation}
\end{mydef}

Instead of knowing the current state of the system,  in some applications, one may also be interested in knowing which initial states the system may start from.
This  is referred to as the initial state estimation problem defined as follows.

\begin{mydef}(Initial-State-Estimate)
Let $G$ be a DES with observable events $\Sigma_o\subseteq \Sigma$ and $\alpha\in P(\mathcal{L}(G))$ be an observed string.
The initial-state-estimate upon the occurrence of $\alpha$, denoted by $\hat{X}_{0,G}(\alpha)$, is the set of initial states the system could start from initially based on observation $\alpha$, i.e.,
\begin{equation}
\hat{X}_{0,G}(\alpha)=\{  x_0\in X_0: \exists  s\in \mathcal{L}(G,x_0)\text{ s.t. }  P(s)=\alpha     \}.
\end{equation}
\end{mydef}

Note that, in the current-state-estimate, we use the observation up to the current instant to estimate the set of all possible states the system can be in at this instant.
If we keep observing more events in the future, our knowledge of the system's state at that instant may be further improved as 
some states in the current-state-estimate may not be consistent with the future observation.
In other words, we can use future information to further improve our knowledge about the system's state for some previous instant.
This is also referred to as the ``smoothing" process; such a smoothed state estimate is also called the delayed-state-estimated defined as follows.

\begin{mydef}(Delayed-State-Estimate)
Let $G$ be a DES with observable events $\Sigma_o\subseteq \Sigma$ and $\alpha\beta\in P(\mathcal{L}(G))$ be an observed string.
The delayed-state-estimate for the instant of $\alpha$ upon the occurrence of $\alpha\beta$, denoted by $\hat{X}_{G}(\alpha \mid \alpha\beta)$, is the set of  states the system could be in $|\beta|$ steps ago when $\alpha\beta$ is observed, i.e.,
\begin{equation}
\hat{X}_{G}(\alpha \mid \alpha\beta)=
\left\{  x\in X:
\begin{array}{c c}
\exists x_0\in X_0,sw\in \mathcal{L}(G,x_0)\text{ s.t. }  \\
P(s)=\alpha\wedge P(sw)=\alpha\beta\wedge   x\in \delta(x_0,s)
\end{array}
\right\}.
\end{equation}
\end{mydef}

\begin{myexm}
Let us consider system $G$ shown in Figure~\ref{fig:G}, where $\Sigma_o=\{a,b,c\}$ and $X_0=\{0,2\}$.
Let us consider observable string $aa\in P(\mathcal{L}(G))$. 
We have $P^{-1}(aa)\cap\mathcal{L}(G)=\{aa,uuaa\}$. 
Then we know that the system may start from   states  $0$ or  $2$, i.e, $\hat{X}_{0,G}(aa)=\{0,2\}$. 
Also, the system may be currently  in states  $4$ or $6$, i.e.,  $\hat{X}_G(aa)=\{4,6\}$. 
From string $aa$, if we further observe event $c$ in the next instant, then we know that $\hat{X}_G(aa\mid aac)=\{6\}$,
since event $c$ cannot occur from state $4$.
\begin{figure}
  \centering
  \subfigure[System $G$ with $\Sigma_o=\{a,b,c\}$]{\label{fig:G}
    \includegraphics[width=0.38\textwidth]{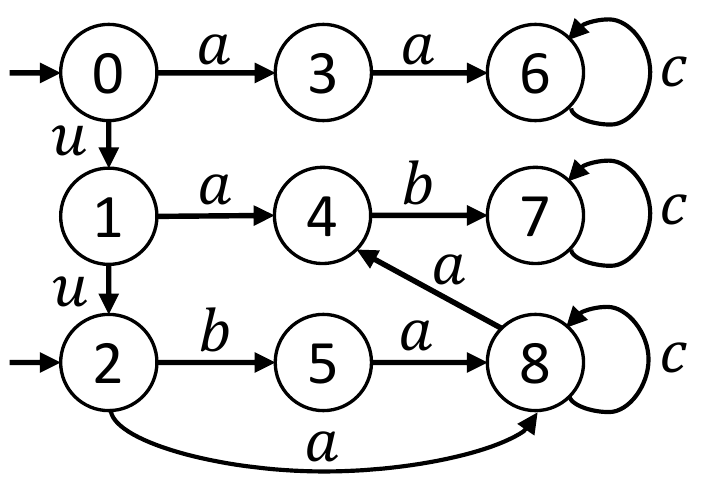}}
  \subfigure[$Obs(G)$]{\label{fig:Obs}
    \includegraphics[width=0.3\textwidth]{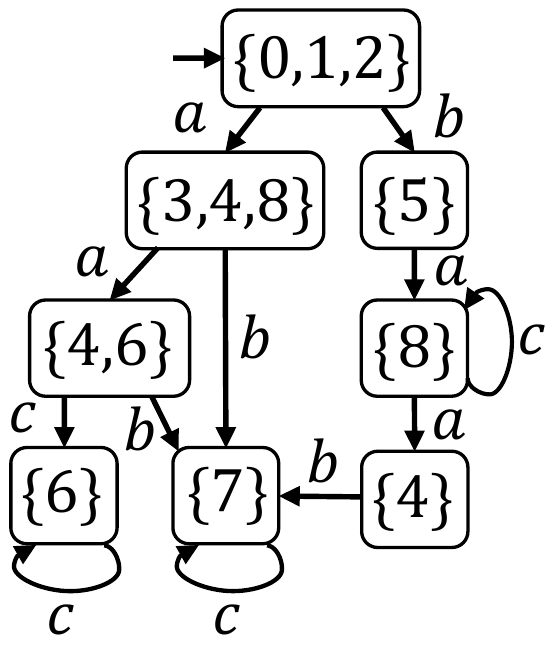}}
 \caption{An arrow labeled with event $\sigma$ from state $x$ to state $x'$ means that $x'\in \delta(x,\sigma)$.
 States with arrows having no predecessor state denote initial states.}
\end{figure}
\end{myexm}

\subsection{State Estimation Techniques}

In this subsection, we provide techniques for  computing different notions of state-estimates.

\subsubsection{Computation of Current-State-Estimate}

The general idea for computing the current-state-estimate is to construct a structure that tracks all possible states consistent with the current observation.
This construction is well-known as the \emph{subset construction} technique  that can be used to convert a NFA to a DFA.
In the DES literature, this structure is usually referred to as the \emph{observer} automaton.

\begin{mydef}(Observer)
Given system $G=(X,\Sigma,\delta,X_0)$ with observable events $\Sigma_o\subseteq \Sigma$,
the observer is a new DFA
\begin{equation}\label{eq:obs}
Obs(G)=(X_{obs},\Sigma_o,\delta_{obs},x_{obs,0}),
\end{equation}
where $X_{obs}\subseteq 2^{X}\setminus \emptyset$ is the set of states,
$x_{obs,0}=\{  x\in  X: \exists x_0\in X_0,w\in \Sigma_{uo}^* \text{ s.t. }x\in \delta(x_0,w) \}$ is the unique initial state and
$\delta_{obs}$ is the deterministic transition function defined by:
for any $q\in X_{obs},\sigma\in \Sigma_o$, we have
\begin{equation}
\delta_{obs}(q,\sigma)=\{ x'\in X: \exists x\in q,\exists w\in \Sigma_{uo}^*\text{ s.t. }x'\in \delta(x,\sigma w)    \}.
\end{equation}
For the sake of simplicity, we will only consider the accessible part of $Obs(G)$.
\end{mydef}

Intuitively, the observer state tracks all possible states the system can be in currently based on the observation.
Formally, we have the following result. 

\begin{myprop}
For any system $G$ with $\Sigma_o\subseteq \Sigma$, its observer $Obs(G)$ has the following properties:
\begin{enumerate}
  \item
  $\mathcal{L}(Obs(G))=P(\mathcal{L}(G))$; and
  \item
  For any $\alpha\in P(\mathcal{L}(G))$, we have $\hat{X}_G(\alpha)=\delta_{obs}(x_{obs,0},\alpha)$.
\end{enumerate}
\end{myprop}

Therefore, given an observation $\alpha\in P(\mathcal{L}(G))$, its current-state-estimate can simply be computed by Algorithm~\ref{alg:cs}.
Note that the complexity of building $Obs(G)$ is $O(|\Sigma|2^{|X|})$, which is exponential in the size of $G$. 
However, we can update the current-state-estimate recursively online and each update step only requires a polynomial complexity.

\begin{algorithm}
\caption{\textsc{Current-State-Estimation}}
\label{alg:cs}
\begin{algorithmic}[1]
\Statex \textbf{\!\!\!\!\!\!Inputs: }   $G$ and $\alpha\in P(\mathcal{L}(G))$
\Statex \textbf{\!\!\!\!\!\!Output: }  $\hat{X}_G( \alpha)$
\State Build $Obs(G)=(X_{obs},\Sigma_o,\delta_{obs},x_{obs,0})$
\State $\hat{X}_G( \alpha)\gets \delta_{obs}(x_{obs,0},\alpha)$
   \State \textbf{return} $\hat{X}_G( \alpha)$
\end{algorithmic}
\end{algorithm}

\subsubsection{Computation of Initial-State-Estimate}

We provide two approaches for the computation of initial-state-estimate.

The first approach is based on the \emph{augmented automaton} that augments the state-space of the original system by tracking where each state starts from.
This approach sometimes is also referred to as the trellis-based approach in the literature.

Formally, given a system $G$, its augmented automaton is a new NFA
\begin{equation}
G_{aug}=(X_{aug}, \Sigma,\delta_{aug},X_{0,aug}  ),
\end{equation}
where
$X_{aug}\subseteq X_0\times X$ is the set of states,
$\delta_{aug}:X_{aug}\times \Sigma\to 2^{X_{aug}}$ is the transition function defined by:
for any $(x_0,x)\in X_{aug},\sigma\in \Sigma$, we have $\delta_{aug}((x_0,x),\sigma)=\{(x_0,x')\in X_0\times X: x'\in \delta(x,\sigma)\}$,
and
$X_{0,aug}=\{(x_0,x_0)\in X_{aug}: x_0\in X_0\}$ is the initial state.

We can see easily that $\mathcal{L}(G)=\mathcal{L}(G_{aug})$ and for each state in $G_{aug}$,
its first component contains its initial state information and its second component contains its current state information.
Let $Obs(G_{aug})=(X_{obs}^{aug},\Sigma_o,\delta_{obs}^{aug},x_{obs,0}^{aug})$ be the observer of $G_{aug}$. 
We can easily show that 
\[
\hat{X}_{0,G}(\alpha)=I_0(  \delta_{obs}^{aug}(x_{obs,0}^{aug},\alpha )    ), 
\]
where for any $q\in X_{obs}^{aug}$, $I_0(q)$ denotes the projection to its first component, i.e., $I_0(q)=\{x_0\in X_0:\exists x\in X\text{ s.t. }(x_0,x)\in q\}$.
This observation suggests Algorithm~\ref{alg:is} for computing the initial-state-estimate.

\begin{algorithm}
\caption{\textsc{Initial-State-Estimation-Aug}}
\label{alg:is}
\begin{algorithmic}[1]
\Statex \textbf{\!\!\!\!\!\!Inputs: }   $G$ and $\alpha\in P(\mathcal{L}(G))$
\Statex \textbf{\!\!\!\!\!\!Output: }  $\hat{X}_{0,G}( \alpha)$
\State Build $G_{aug}$
\State Build $Obs(G_{aug})=(X_{obs}^{aug},\Sigma_o,\delta_{obs}^{aug},x_{obs,0}^{aug})$
\State $\hat{X}_{0,G}( \alpha)\gets I_0(  \delta_{obs}^{aug}(x_{obs,0}^{aug},\alpha )    )$
   \State \textbf{return} $\hat{X}_{0,G}( \alpha)$
\end{algorithmic}
\end{algorithm}

The idea of the augmented-automaton-based approach is also very useful for many other purposes, e.g., control synthesis for initial state estimation,
as it only uses the dynamic of the system up to the current point.
However, the complexity of Algorithm~\ref{alg:is} is $O(|\Sigma|2^{|X|^2})$ as the size of $G_{aug}$ is quadratic in the size of $G$.
Here we  provide the second approach for computing the initial-state-estimate with a lower complexity based on the \emph{reversed automaton} of $G$.

For any NFA $G=(X,\Sigma,\delta,X_0)$, its  reversed automaton is  a new NFA
\begin{equation}
G_R=(X,\Sigma,\delta_R,X),
\end{equation}
where the transition function $\delta_R: X\times \Sigma\to 2^X$ is defined by:
$\forall x,x'\in X,\sigma\in\Sigma: x'\in \delta_R(x,\sigma)\Leftrightarrow x\in \delta(x',\sigma)$.
Note that  the initial state of $G_R$ is the entire state space.
For any string $s=\sigma_1\sigma_2\dots\sigma_n \in \Sigma^*$, we denote by $s_R$ its reversed string, i.e., $s_R=\sigma_{n}\dots\sigma_2\sigma_1$.
Then  the following result shows that the initial-state-estimate of a string in $G$ can be computed based on 
the current-state-estimate of its reversed string in the reversed automaton $G_R$.

\begin{myprop}\cite{wu2013comparative}
For any $\alpha\in P(\mathcal{L}(G))$, we have $\hat{X}_{0,G}(\alpha)=\hat{X}_{G_R}(\alpha_R)\cap X_0$.
\end{myprop}

Based on the above result,  Algorithm~\ref{alg:is-2} is proposed  to compute the initial-state-estimate
and its complexity is only  $O(|\Sigma|2^{|X|})$.
\begin{algorithm}
\caption{\textsc{Initial-State-Estimation-Rev}}
\label{alg:is-2}
\begin{algorithmic}[1]
\Statex \textbf{\!\!\!\!\!\!Inputs: }   $G$ and $\alpha\in P(\mathcal{L}(G))$
\Statex \textbf{\!\!\!\!\!\!Output: }  $\hat{X}_{0,G}( \alpha)$
\State Build $G_R$
\State Build $Obs(G_R)=(X_{obs}^{R},\Sigma_o,\delta_{obs}^{R}, X)$
\State $\hat{X}_{G_R}(\alpha_R)\gets   \delta_{obs}^{R}(X,\alpha_R)$
   \State \textbf{return} $\hat{X}_{0,G}( \alpha)\gets \hat{X}_{G_R}(\alpha_R)\cap X_0$
\end{algorithmic}
\end{algorithm}

\subsubsection{Computation of Delayed-State-Estimate}

For any $\alpha\beta\in P(\mathcal{L}(G))$, the delayed-state-estimate $\hat{X}_G(\alpha\mid \alpha \beta)$ involves two parts of information:
the current information $\alpha$ and the future information $\beta$.
Recall that we are interested in estimating states for the instant of $\alpha$.
The following result shows that the delayed-state-estimate can simply be separated as two parts that do not depend on each other.

\begin{myprop}\label{prop:delay}\cite{yin2017new}
For any $\alpha\beta\in P(\mathcal{L}(G))$, we have
\[
\hat{X}_G(\alpha\mid \alpha \beta)=  \hat{X}_{G}(\alpha)\cap \hat{X}_{0,G}(\beta)=  \hat{X}_{G}(\alpha)\cap \hat{X}_{G_R}(\beta_R).
\]
\end{myprop}

Based on Proposition~\ref{prop:delay},   Algorithm~\ref{alg:dse} can be used to compute the delayed-state-estimate and its complexity is also  $O(|\Sigma|2^{|X|})$.
\begin{algorithm}
\caption{\textsc{Delayed-State-Estimation}}
\label{alg:dse}
\begin{algorithmic}[1]
\Statex \textbf{\!\!\!\!\!\!Inputs: }   $G$ and $\alpha\beta\in P(\mathcal{L}(G))$
\Statex \textbf{\!\!\!\!\!\!Output: }  $\hat{X}_{G}( \alpha\mid \alpha\beta)$
\State Build $Obs(G)=(X_{obs},\Sigma_o,\delta_{obs},x_{obs,0})$
\State $\hat{X}_G( \alpha)\gets \delta_{obs}(x_{obs,0},\alpha)$
\State Build $G_R$
\State Build $Obs(G_R)=(X_{obs}^{R},\Sigma_o,\delta_{obs}^{R}, X)$
\State $\hat{X}_{G_R}(\beta_R)\gets   \delta_{obs}^{R}(X,\beta_R)$
   \State \textbf{return} $\hat{X}_G(\alpha\mid \alpha \beta)\gets  \hat{X}_{G}(\alpha)\cap \hat{X}_{G_R}(\beta_R)$
\end{algorithmic}
\end{algorithm}

\begin{myexm}
We still consider system $G$ shown in Figure~\ref{fig:G}, where $\Sigma_o=\{a,b,c\}$ and $X_0=\{0,2\}$.
Its observer $Obs(G)$ is shown in Figure~\ref{fig:Obs}.
For example, string $aa$ reaches state $\{4,6\}$ from the initial state in $Obs(G)$.
Therefore, we have $\hat{X}_G(aa)=\{4,6\}$.
To preform   initial state estimation and delayed state estimation,
we need to build the reversed automaton $G_R$ and  its observer $Obs(G_R)$  shown in Figures~\ref{fig:GR} and~\ref{fig:ObsR}, respectively.
For example, string $cba$ reaches state $\{0,1,8\}$ from the initial state in $Obs(G_R)$.
Therefore, we have $\hat{X}_{0,G}(abc)=\hat{X}_{G_R}((abc)_R)\cap X_0=\{0,1,8\}\cap \{0,2\}=\{0\}$,
i.e., we know for sure that the system was initially from state $0$.
To computed the delayed-state-estimate $\hat{X}_{G}(a \mid aac)$,
we have
$\hat{X}_{G}(a \mid aac)=\hat{X}_{G}(a)\cap \hat{X}_{G_R}(ca)=\{3,4,8\}\cap\{0,1,2,3,5\}=\{3\}$.
\begin{figure}
  \centering
  \subfigure[$G_R$ with $\Sigma_o=\{a,b,c\}$]{\label{fig:GR}
    \includegraphics[width=0.32\textwidth]{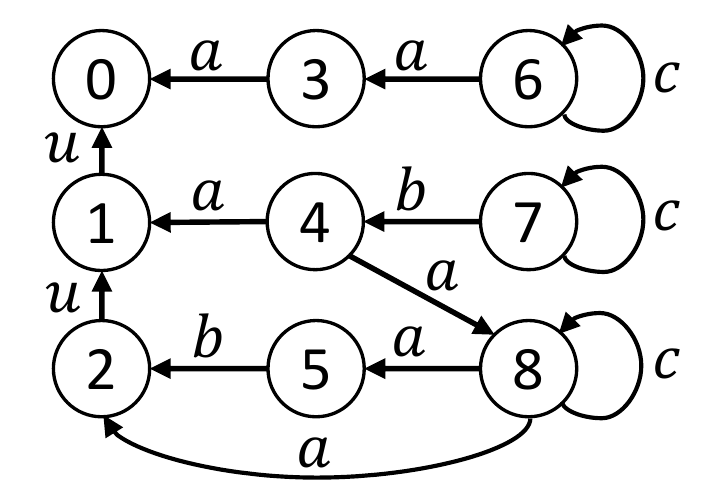}}
  \subfigure[$Obs(G_R)$]{\label{fig:ObsR}
    \includegraphics[width=0.53\textwidth]{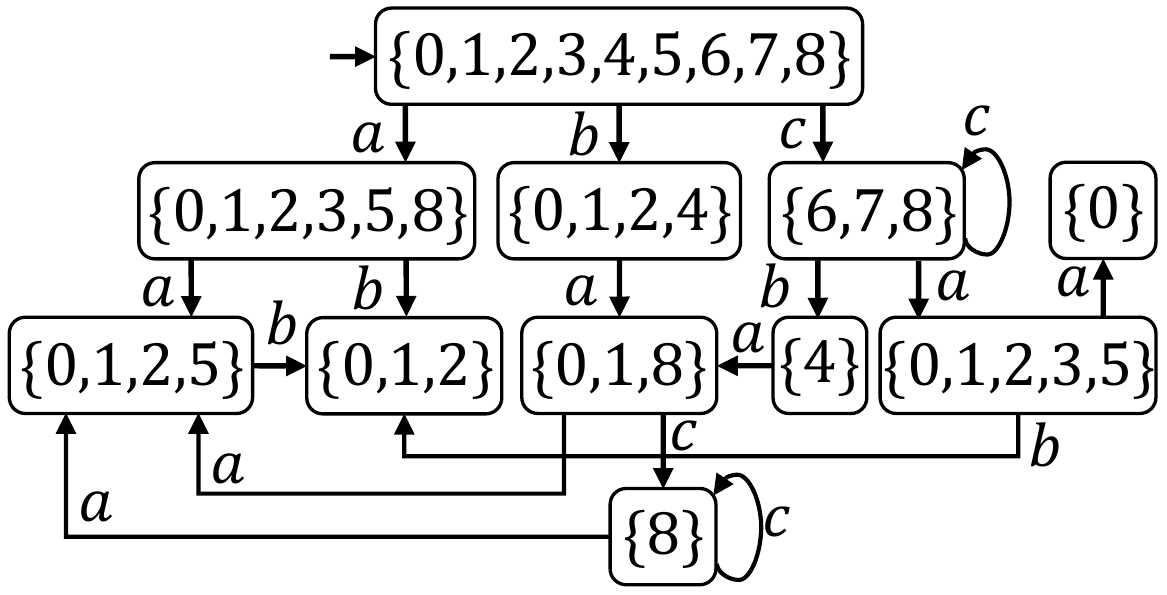}}
 \caption{Initial state estimation using the reversed automaton. Note that all states in $G_R$ are initial.}
\end{figure}
\end{myexm}

\section{Properties of Partially-Observed DES}\label{sec:4}
In this section, we discuss several important properties in partially-observed DES.
As we mentioned before, we will only focus on observational properties.

\subsection{Detectability}

In the previous section, we have provided algorithms for computing state estimates.
Then  the natural question arises as to can the state estimation algorithm   (eventually) provides precise state information.
This is referred to as the \emph{detectability verification problem}.
Here, we consider three most fundamental types of detectability.

\begin{mydef}(Detectability  \cite{shu2007detectability,shu2013detectability,shu2013delayed} )
Let  $G$ be a DES with $\Sigma_o\subseteq \Sigma$. Then $G$ is said to be
\begin{itemize}
  \item
  current-state detectable if $(\exists n \in  \mathbb{N})(\forall \alpha \in  P(\mathcal{L}(G)): |\alpha| \geq  n)[|\hat{X}_G(\alpha)|=1]$;
  \item
  initial-state detectable if $(\exists n \in  \mathbb{N})(\forall \alpha \in  P(\mathcal{L}(G)): |\alpha| \geq  n)[ |\hat{X}_{0,G}(\alpha)|=1]$;
  \item
  delayed detectable (w.r.t.\ parameters $k_1,k_2\!\in\! \mathbb{N}$)  if $(\forall \alpha\beta\!\in\! P(\mathcal{L}(G)): |\alpha|\geq k_1,|\beta|\geq k_2)[  |\hat{X}_{G}(\alpha \mid \alpha\beta)|=1]$.
\end{itemize}
\end{mydef}

Intuitively,  current-state detectability (respectively, initial-state detectability) requires that the current state (initial state) of the system can always be detected unambiguously within a finite delay. Note that, for initial-state-estimate, once the initial state is detected, we know it for sure forever, which is not the case for current-state-estimation.
Delayed detectability requires that, for any specific instant after $k_1$ steps, we can always unambiguously determine the precise state of the system at that instant 
with at most $k_2$ steps of information delay.
That is, one is allowed to use future information to ``smooth" the state estimate of a previous instant.

The concept of detectability was initially  proposed by Shu and Lin in \cite{shu2007detectability}, which generalizes the concept of observability studied in \cite{ozveren1990observability}.
There are also other notions of detectability proposed in the literature.
For example, weak detectability \cite{shu2007detectability} requires that the state of the system can be detected for \emph{some} path generated by the system.
Also, \cite{shu2007detectability} proposed the notion of periodic detectability, which requires that  the state of the system can be detected periodically.
In \cite{hadjicostis2016k}, the authors proposed the notion of $K$-detectability by replacing the detection condition $|\hat{X}_G(\alpha)|=1$ as $|\hat{X}_G(\alpha)|\leq K$,
where $K\in \mathbb{N}$ is a positive integer specifying the detection precision.
A generalized version of detectability was proposed in \cite{shu2011generalized}.
The reader is referred to \cite{sasi2018detectability,keroglou2017verification,masopust2019complexity,masopust2018complexity,shu2008state,yin2017verification,zhang2017problem}  for more references on detectability.

\subsection{Diagnosability and Prognosability of Fault}

Another important application of partially-observed DES is the fault diagnosis/prognosis problem.
In this setting, we assume that system $G$ may have some fault modeled as fault events  $\Sigma_F\subseteq \Sigma$.
For any string $s\in \Sigma^*$, we write $\Sigma_F\in s$ if $s$ contains a fault event in $\Sigma_F$.
We define $\Psi(\Sigma_F):=\{s e_f\in \mathcal{L}(G): e_f\in \Sigma_{F}\}$ as the set of strings that end up with fault events.
In the fault diagnosis problem, we assume that all fault events are unobervable; otherwise, it can be diagnosed trivially.
Without loss of generality, we can further assume that the state space of $G$ is partitioned as fault states and non-fault states
\begin{equation}
X=X_N\dot{\cup}X_F, 
\end{equation}
such that
\begin{itemize}
  \item
  $\forall x_0\in X_0,\forall s\in \mathcal{L}(G,x_0):  \Sigma_F\notin s\Rightarrow \delta(x_0,s)\subseteq X_N$; and
  \item
  $\forall x_0\in X_0,\forall s\in \mathcal{L}(G,x_0):  \Sigma_F\in s\Rightarrow \delta(x_0,s)\subseteq X_F$.
\end{itemize}
This assumption can be fulfilled by taking the product   between $G$ and a new automaton with two states capturing the occurrence of fault;
see, e.g., \cite{Lbook}.

To diagnose the occurrence of fault, the current state estimation technique can be applied.
Specifically,  for any observation  $\alpha\in P(\mathcal{L}(G))$, we know that
\begin{itemize}
  \item
  the fault has occurred for sure if $\hat{X}_G(\alpha)\subseteq X_F$;
  \item
  the fault has not occurred for sure if $\hat{X}_G(\alpha)\subseteq X_N$;
  \item
  the fault may have occurred but it is uncertain  if $\hat{X}_G(\alpha)\cap X_N\not=\emptyset$ and $\hat{X}_G(\alpha)\cap X_F\not=\emptyset$.
\end{itemize}
Then  the natural question arises as to can we always determine the occurrence of fault  within a finite number of delays.
This is captured by the notion of diagnosability as follows. 

\begin{mydef}(Diagnosability)
System $G$ is said to be diagnosable w.r.t.\ $\Sigma_o$ and $\Sigma_F$ if
$(\forall s\in \Psi(\Sigma_F))(\exists n\in \mathbb{N}) (\forall t\in \mathcal{L}(G)/s:|t|\geq n)
[ \hat{X}_G(P(st))\subseteq X_F ]$.
\end{mydef}

\begin{remark}
For the sake of simplicity, our definition of diagnosability is based on the current-state-estimate and the pre-specified  fault states.
Diagnosability was originally defined by \cite{sampath1995diagnosability} purely based on   languages as follows
\begin{align}
 (\forall s\!\in\! \Psi(\Sigma_F))(\exists n\!\in\!\mathbb{N})(\forall t\!\in \!\mathcal{L}(G)/s:|t|\geq n)
(\forall w\!\in\! P^{-1}(P(st)) \cap \mathcal{L}(G)) [\Sigma_F\!\in\! w]\nonumber
 \end{align}
 One can easily check that the language-based definition and the current-state-estimate-based definition are equivalent. 
 Also, in general,   the non-fault behavior can be described as a   specification language rather than fault events; 
this formulation can also be transformed to our event-based fault setting by refining the state-space of the system; see, e.g., \cite{yoo2008diagnosis}.
\end{remark}

The concept of diagnosability of DES was first introduced in \cite{lin1994diagnosability}, where state-based faults are considered.
In \cite{sampath1995diagnosability}, the authors introduced the language-based formulation of diagnosability.
Since then  many variations of diagnosability have been studied in the literature.
For example, model reduction for diagnosability was studied in \cite{zad2003fault}.
Diagnosability of repeated/intermittent faults was studied in \cite{jiang2003diagnosis,contant2004diagnosis,fabre2018diagnosability}. 
The reader is referred to the recent survey \cite{zaytoon2013overview} for more references on diagnosability analysis.

In some applications,  we may want to \emph{predict} the occurrence of fault before it actually occurs.
This problem is referred to as the \emph{fault prognosis problem}.
Still, we assume that $\Sigma_F\subseteq\Sigma$ is the set of fault events and $X_N\subseteq X$ is the set of non-fault states.
However, in the fault prognosis problem, a fault event needs not be unobservable.
To formulate the problem, we define the  following two sets of states:
\begin{itemize}
  \item
  boundary states, $\partial(G)=\{x\in X_N: \exists e_f\in \Sigma_F\text{ s.t. }\delta(x,e_f)!\}$; and
  \item
  indicator states,
  $\Im(G)=\{x\in X_N: \exists n\in \mathbb{N},\forall s\in \mathcal{L}(G,x)\text{ s.t. }|s|>n\Rightarrow \Sigma_F \!\in\! s\}$.
\end{itemize}
Intuitively, a boundary state is a non-fault state  from which a fault event \emph{may} occur in the next step,
and  an indicator state is a state from which a fault event  will occur \emph{for sure} within a finite number of steps.
With the help indicator states, we can also use current-state-estimation algorithm to perform online fault progsnosis.
Specifically,  for any observation  $\alpha\in P(\mathcal{L}(G))$, we know that
\begin{itemize}
  \item
  the fault will occur for sure in a finite number of steps if $\hat{X}_G(\alpha)\subseteq \Im(G)$;
  \item
  the fault is not guaranteed to occur within any finite number of steps  if $\hat{X}_G(\alpha)\not\subseteq \Im(G)$.
\end{itemize}
Similarly,  the natural question arises as to can we successfully predict the occurrence of fault in the sense that:
(i) there is no missed fault; and (ii) there is no false alarm.
This is captured by the notion of prognosability as follows.

\begin{mydef}(Prognosability)
System $G$ is said to be prognosable w.r.t.\ $\Sigma_o$ and $\Sigma_F$ if
$(\forall s\in \mathcal{L}(G):  \delta(s)\cap\partial(G) \neq\emptyset )
(\exists t\in \overline{\{s\}})
[ \hat{X}_G(P(t))\subseteq\Im(G)]$.
\end{mydef}

Prognosability is also referred to as predictability in the literature.
Intuitively, prognosability requires that, for any string that reaches a boundary state, it has a prefix for which we can claim unambiguously that fault will occur for sure in the future,
i.e., we can issue a fault alarm.
Still, for the sake of simplicity, here we characterize prognosability using current-state-estimate, boundary states and indicator states.
Our definition is also  equivalent to the language-based definition in  \cite{jeron2008predictability,genc2009predictability}, where predictability was originally introduced.
Prognosability also has several variations in the literature.
For example, in \cite{yin2016decentralized}, two performance bounds were proposed to characterize how early a fault alarm can be issued 
and when a fault is guaranteed to occur once an alarm is issued.

\subsection{State Disambiguation and Observability}

In some applications, the purpose of state estimation is to distinguish some states.
This problem is referred to as the \emph{state disambiguation problem} \cite{wang2007algorithm,sears2014computing,yin2018minimization}.
Formally, the specification of this problem is defined as a set of state pairs $T_{spec}\subseteq X\times X$
and we want to make sure that we can always distinguish between a pair of states in $T_{spec}$.

\begin{mydef}(Distinguishability)
System $G$ is said to be distinguishable w.r.t.\ $\Sigma_o$ and $T_{spec}\subseteq X\times X$ if
$(\forall \alpha\in P(\mathcal{L}(G)))  [ \left(\hat{X}_G(\alpha)\times \hat{X}_G(\alpha) \right)  \cap T_{spec}= \emptyset]$.
\end{mydef}

Distinguishability is very useful as many important properties in the literature can be formulated as distinguishability.
For example,  one can check that prognosability can actually be re-written as  ditinguishability with specification $T_{spec}=   \partial(G) \times  (X_N\setminus \Im(G))$.
Observability is another important property in partially-observed DES, which together with controllability provide the necessary and sufficient conditions for the existence of a supervisor achieving a desired language. The reader is referred to \cite{lin1988observability,cieslak1988} for formal definition of observability.
This property is also a special case of distinguishability (possibly after state-space refinement) as it essentially requires that we can distinguish two states at which different control actions are needed; see, e.g.,  \cite{wang2007algorithm}.

\subsection{Opacity}

Finally, state estimation is also useful in information-flow security analysis, which is an important topic in cyber-physical systems.
In this setting, we assume that the system is also monitored by a passive intruder (eavesdropper) that can observe the occurrences of events in $\Sigma_o$.
Furthermore, we assume that the system has a ``secret" that does not want to be revealed to the intruder.
In general, what is a secret is problem dependent.
Here, we consider a simple scenario where the secret is modeled as a set of secret states $X_S\subseteq X$.
Then we use the notion of opacity to characterize whether or not the secret can be revealed to the intruder.

\begin{mydef}(Opacity \cite{saboori2012verification,saboori2013verification,wu2013comparative,yin2017new})
Let  $G$ be a DES with $\Sigma_o\subseteq \Sigma$ and secret states $X_S\subseteq X$. Then $G$ is said to be
\begin{itemize}
  \item
  current-state opaque if, for any  $\alpha\in P(\mathcal{L}(G))$, we have  $\hat{X}_G(\alpha) \not\subseteq X_S$;
  \item
  initial-state opaque if, for any  $\alpha\in P(\mathcal{L}(G))$, we have  $\hat{X}_{0,G}(\alpha) \not\subseteq X_S$;
  \item
  infinite-step opaque  if, for any   $\alpha\beta\in P(\mathcal{L}(G))$, we have  $ \hat{X}_G(\alpha \mid \alpha\beta) \not\subseteq X_S$.
\end{itemize}
\end{mydef}

Essentially, opacity is a confidentiality property  capturing the plausible deniability of the system's ``secret" in the presence of an outside observer that is potentially malicious.
More specifically, current-state opacity (respectively, initial-state opacity) requires that
the intruder should never know for sure that the system is currently at (respectively, initially from) a secret state.
The system is said to be infinite-step opaque if the intruder can never determine for sure that the system was at a secret state for any specific instant even based on the future information.

Opacity was originally introduced in the computer science literature \cite{mazare2004using}.
Then it was introduced to the framework of DES by \cite{badouel2007concurrent,bryans2008opacity,saboori2007notions}.
There are also many variations of opacity studied in the literature.
For example, in \cite{lin2011opacity}, Lin formulated  opacity using a language-based framework, where the notions of strong opacity and weak opacity are proposed.
In \cite{wu2013comparative}, a concept called initial-and-final-state opacity was provided; the authors also studied the transformations among several notions of opacity.
When one is only allowed to used a bounded delayed information to improve the state estimate for a previous instant (or we do not care about the secret anymore after some delays),
infinite-step opacity becomes to $K$-step opacity \cite{saboori2011verification}, where $K$ is a  non-negative integer capturing the delay bound.
Opacity is also closely related to another two information-flow security properties called anonymity \cite{sweeney2002k} and non-interference \cite{hadj2005verification,hadj2005characterizing}.
The reader is referred to the survey \cite{jacob2016overview}  for more references on opacity.

\section{Verification Techniques}\label{sec:5}

In this section, we provide techniques for verifying observational properties introduced in the previous section.
First, we show that most of the properties in  partially-observed DES can be verified using the observer structure
and 
some important properties can be more efficienty verified using the twin-plant technique in polynomial-time.

\subsection{Observer-Based Verification}

\subsubsection{Verification of Detectability}
First, we study the verification of current-state detectability.
Recall that, current-state detectability requires that  $(\exists n\in \mathbb{N})(\forall \alpha\in P(\mathcal{L}(G)): |\alpha|\geq n)[  |\hat{X}_G(\alpha)|=1]$.
In other words, a system is \emph{not} current-state detectable if
there is an arbitrarily long observation string such that the current-state-estimate is \emph{not always} a singleton starting from any instant.
Since $P(\mathcal{L}(G))$ is a regular language, due to the Pumping lemma, the existence of such an arbitrarily long string is equivalent to the existence of a cycle in
$Obs(G)$ in which a state is not a singleton.
This immediately suggests Algorithm~\ref{alg:dect} for the verification of current-state detectability.
Initial-state detectability can also be checked in the same manner using $G_{aug}$.
The only differences from  Algorithm~\ref{alg:dect}  are
 (i) we  need to consider $Obs(G_{aug})$ rather than $Obs(G)$ in line~1; and
(ii) in line~2, we need to check the existence of a cycle $q_0q_1\dots q_n$ in $Obs(G_{aug})$ such that $|I_0(q_i)|>1$ for some $i=1,\dots,n$.

\begin{algorithm}
\caption{\textsc{Cur-State-Dect-Ver-Obs}}
\label{alg:dect}
\begin{algorithmic}[1]
\Statex \textbf{\!\!\!\!\!\!Inputs: }   $G$
\Statex \textbf{\!\!\!\!\!\!Output: } Current-State Detectable or Not
\State Build $Obs(G)=(X_{obs},\Sigma_o,\delta_{obs},x_{obs,0})$
\If {there exists a cycle $q_0q_1\dots q_n$ in $Obs(G)$ such that $|q_i|>1$ for some $i=1,\dots,n$}
\State \textbf{return} $G$ is not current-state detectable
\Else
 \State \textbf{return} $G$ is current-state detectable
\EndIf
\end{algorithmic}
\end{algorithm}

To check delayed detectability (for parameters $k_1,k_2\in \mathbb{N}$),  first we recall that
the delayed-state-estimate can be computed by $\hat{X}_{G}(\alpha\mid \alpha\beta)=\hat{X}_{G}(\alpha)\cap \hat{X}_{G_R}(\beta_R)$.
Therefore, delayed detectability can be checked by the following steps:
\begin{itemize}
  \item
  First, we compute all possible current-state-estimate $\hat{X}_{G}(\alpha)$ reachable via some string $\alpha$ whose length is greater than or equal to $k_1$;
  \item
  Then, we compute all possible current-state-estimate of $\hat{X}_{G_R}(\beta_R)$ in $G_R$ reachable via some string $\beta_R$ whose length is greater than or equal to $k_2$;
  \item
  Finally, we test whether or not there exist such $\hat{X}_{G}(\alpha)$ and  $\hat{X}_{G_R}(\beta_R)$ such that $|\hat{X}_{G}(\alpha)\cap \hat{X}_{G_R}(\beta_R)|>1$.
  If so, then it means that some instant after $k_1$ steps cannot be determined within $k_2$ steps of delay, i.e,  the system is not delayed detectable.
\end{itemize}
This procedure is formalized by  Algorithm~\ref{alg:ds}, where states in lines~3 and~4 can be computed by a simple depth-first search or a breath-first search. 

\begin{algorithm}
\caption{\textsc{Delay-State-Dect-Ver-Obs}}
\label{alg:ds}
\begin{algorithmic}[1]
\Statex \textbf{\!\!\!\!\!\!Inputs: }   $G$
\Statex \textbf{\!\!\!\!\!\!Output: } Delayed Detectable or Not
\State Build $Obs(G)=(X_{obs},\Sigma_o,\delta_{obs},x_{obs,0})$
\State Build $Obs(G_R)=(X_{obs}^R,\Sigma_o,\delta_{obs}^R,X)$
\ForAll{$q_1\in X_{obs}$ that be can reached by a string longer than $k_1$ in $Obs(G)$}
\ForAll{$q_2\in X_{obs}^R$ that be can reached by a string longer than $k_2$ in $Obs(G_R)$}
\If{$|q_1\cap q_2|>1$}
  \State  \textbf{return} $G$ is not delayed detectable
 \EndIf
\EndFor
\EndFor
  \State  \textbf{return} $G$ is delayed detectable
\end{algorithmic}
\end{algorithm}

\subsubsection{Verification of Diagnosability}

The verification of diagnosability is very similar to the case of current-state detectability.
Specifically, a system is \emph{not} diagnosable if there is an arbitrarily long string \emph{after the occurrence of fault} such that the current-state-estimate is \emph{not} a subset of $X_N$
starting from any instant.
Therefore, the general idea is to replaced condition $|q_i|>1$ in Algorithm~\ref{alg:dect} by $q_i\not\subseteq X_F$.
However, we need to do a little bit more here since we are only interested in   arbitrarily long uncertain strings \emph{after the occurrence of fault};
this is also referred to as an \emph{indeterminate cycles} in the literature \cite{sampath1995diagnosability}.
In other words, the existence of an arbitrarily long uncertain but non-fault string does not necessarily violate diagnosability.

To this end, we need to compose $Obs(G)$ with the dynamic of the original system $G$.
Specifically, let $Obs(G)$ be the observer of $G$. We define
\begin{equation}
\widetilde{Obs}(G)=(X_{obs},\Sigma,\tilde{\delta}_{obs},x_{obs,0})
\end{equation}
as the  DFA by adding self-loops of all unobservable events at each state in $X_{obs}$.
One can easily check that $\mathcal{L}(G)=\mathcal{L}(G\times \widetilde{Obs}(G))$. 
Then, by looking at the first component of each state in $G\times \widetilde{Obs}(G)$, we know whether a fault event has occurred or not. 
Therefore, we can check diagnosability based on $G\times \widetilde{Obs}(G)$ by Algorithm~\ref{alg:dia-obs}.

\begin{algorithm}
\caption{\textsc{Diag-Ver-Obs}}
\label{alg:dia-obs}
\begin{algorithmic}[1]
\Statex \textbf{\!\!\!\!\!\!Inputs: }   $G$ and $\Sigma_F$
\Statex \textbf{\!\!\!\!\!\!Output: }  Diagnosable or Not
\State Build $Obs(G)=(X_{obs},\Sigma_o,\delta_{obs},x_{obs,0})$
\State Augment $Obs(G)$ as $\widetilde{Obs}(G)=(X_{obs},\Sigma,\tilde{\delta}_{obs},x_{obs,0})$ by adding unobservable self-loops
\State Compute $G\times \widetilde{Obs}(G)$
\If {there exists a cycle $(x_0,q_0)(x_1,q_1)\dots (x_n,q_n)$ in $G\times \widetilde{Obs}(G)$ such that
$x_i\in X_F$ and $q_i\not\subseteq X_F$ for some $i=1,\dots,n$}
\State \textbf{return} $G$ is not diagnosable
\Else
 \State \textbf{return} $G$ is  diagnosable
\EndIf
\end{algorithmic}
\end{algorithm}

\subsubsection{Verification of Distinguishability and Opacity}

The verification of distinguishability is straightforward using the observer.
Specifically, we just need to check whether or not there exists a state $q\in X_{obs}$  in $Obs(G)$ such that $(q\times q)  \cap T_{spec}\not= \emptyset$.
If so, the system is not distinguishable; otherwise, it is distinguishable.
Since we have discussed that prognosability is a special case of distinguishability, it can also be checked in the same manner.

Similarly, current-state opacity can be verified by  checking whether or not there exists a state $q\in X_{obs}$ in $Obs(G)$ such that $q\subseteq X_S$.
If so, the system is not opaque; otherwise, it is opaque.
To check initial-state opacity, we need to construct $Obs(G_R)$ and to check  whether or not there exists a state $q\in X_{obs}^R$ in $Obs(G_R)$ such that $q\cap X_0  \subseteq X_S$.

The verification of infinite-step opacity is similar to the case of delayed detectability; both involve delayed-state-estimate that can be computed according to Proposition~\ref{prop:delay}.
The verification procedure for infinite-step opacity is provided in  Algorithm~\ref{alg:if} with complexity  $O(|\Sigma|4^{|X|} )$.

\begin{algorithm}
\caption{\textsc{Inf-Opa-Ver-Obs}}
\label{alg:if}
\begin{algorithmic}[1]
\Statex \textbf{\!\!\!\!\!\!Inputs: }   $G$
\Statex \textbf{\!\!\!\!\!\!Output: } Infinite-Step Opaque or Not
\State Build $Obs(G)=(X_{obs},\Sigma_o,\delta_{obs},x_{obs,0})$
\State Build $Obs(G_R)=(X_{obs}^R,\Sigma_o,\delta_{obs}^R,X)$
\ForAll{$q_1\in X_{obs}$}
\ForAll{$q_2\in X_{obs}^R$}
\If{$\emptyset\neq q_1\cap q_2 \subseteq X_S$}
  \State  \textbf{return} $G$ is not infinite-step opaque
 \EndIf
\EndFor
\EndFor
  \State  \textbf{return} $G$  is  infinite-step opaque
\end{algorithmic}
\end{algorithm}

\subsection{Twin-Plant-Based Verification}
In the previous section, we have shown that the observer  can be used for the verification of all properties introduced.
However, the size of the observer is exponential in the size of the system.
Then  the natural question arises as to can we find polynomial-time algorithms for verifying these properties.
Unfortunately, it has been shown in \cite{cassez2012synthesis} that deciding opacity is PSPACE-complete; hence, no polynomial-time algorithm exists.
However, it is indeed possible to check diagnosability, detectability and distinguishability in polynomial-time by using the \emph{twin-plant} structure.
This structure was originally proposed in \cite{tsitsiklis1989control} for the verification of observability;
later on it has been used for verifying diagnosability (under the name of  ``verifier") by \cite{jiang2001polynomial,yoo2002polynomial} and for verifying detectability (under the name of ``detector") by \cite{shu2010detectability}.

\begin{mydef}(Twin-Plant)
Given system $G=(X,\Sigma,\delta,X_0)$ with observable events $\Sigma_o\subseteq \Sigma$,
the twin-plant is a new  NFA
$V(G)=(X_V,  \Sigma_V  ,\delta_V  ,X_{0,V})$, where
\begin{itemize}
  \item
  $X_V\subseteq X\times X$ is the set of states;
  \item
  $\Sigma_V=(\Sigma_o\times\Sigma_o)\cup (\Sigma_{uo}\times \{\epsilon\})\cup (\{\epsilon\}\times \Sigma_{uo})$  is the set of events;
  \item
  $X_{0,V}=X_0\times X_0$ is the set of initial states;
  \item
  $\delta_V:X_V\times\Sigma_V\rightarrow 2^{X_V}$ is the partial transition function defined by:
  for any state $(x_1,x_2) \in X_V$ and event $\sigma\in \Sigma$
\end{itemize}
(a) If $\sigma\in \Sigma_o$, then the following transition is defined
\begin{equation}
\delta_V((x_1,x_2),(\sigma,\sigma))= \delta(x_1,\sigma)\times \delta(x_2,\sigma)
\end{equation}
(a) If $\sigma\in \Sigma_{uo}$, then the following transitions are defined
\begin{align}
\delta_V((x_1,x_2),(\sigma,\epsilon))&= \delta(x_1,\sigma)\times \{x_2\}\\
\delta_V((x_1,x_2),(\epsilon,\sigma))&=\{x_1\}\times \delta(x_2,\sigma)
\end{align}
Hereafter, we only consider the accessible part of $V(G)$.
\end{mydef}

Intuitively, the twin-plant $V$ tracks all pairs of observation equivalent strings in $G$.
Specifically, if $s_1,s_2\in \mathcal{L}(G)$ are two strings in $G$ such that $P(s_1)=P(s_2)$,
then there exists a string $s\in \mathcal{L}(V(G))$ in $V(G)$ such that its first and second components are $s_1$ and $s_2$, respectively.
On the other hand, for any string $s=(s_1,s_2)\in \mathcal{L}(V(G))$ in $V(G)$, we have that $P(s_1)=P(s_2)$.

The twin-plant can be applied directly for the verification of distinguishability.
Specifically, we need to check if $V(G)$ contains a pair of states in the specification.
This procedure is presented in Algorithm~\ref{alg:dist}.

According to the definition of detectability, we know that the system is not detectable if and only if there exist two arbitrarily long strings having the same observation,
such that these two strings lead to two different states.
As we discussed above, all such string pairs can be captured by the twin-plant.
This suggests Algorithm~\ref{alg:dect-tp} for the verification of current-state detectability.

The case of diagnosability is similar.
Specifically, a system is not diagnosable  if there exist an arbitrarily long fault string and a non-fault string such that they have the same observation.
This condition can be checked by Algorithm~\ref{alg:dia-tp}.
Note that, we have already assumed that there is no unobservable cycle in $G$.
Otherwise, we need to add the following condition to the ``if condition" in line~2 of Algorithm~\ref{alg:dia-tp} to obtain an arbitrarily long fault string:
\[\exists j=0,\dots,n-1,\exists \sigma_V\notin \Sigma_{uo}\times \{\epsilon\}: (x_{j+1}^1,x_{j+1}^2)\in \delta_V((x_j^1,x_j^2),\sigma_V).\]
Since the size of the twin-plant is only quadratic in the size of $G$,
distinguishability, detectability and diagnosability can all be checked in polynomial-time,
which is better than the observer-based approach.

\begin{algorithm}
\caption{\textsc{Dist-Ver-Obs}}
\label{alg:dist}
\begin{algorithmic}[1]
\Statex \textbf{\!\!\!\!\!\!Inputs: }   $G$ and $T_{spec}$
\Statex \textbf{\!\!\!\!\!\!Output: }  Distinguishable or Not
\State Build $V(G)=(X_V,  \Sigma_V  ,\delta_V  ,X_{0,V})$
\If {there exists a state $(x_1,x_2)\in X_V$ in $V(G)$ such that  $(x_1,x_2)\in T_{spec}$}
\State \textbf{return} $G$ is not distinguishable
\Else
 \State \textbf{return} $G$ is distinguishable
\EndIf
\end{algorithmic}
\end{algorithm}

\begin{algorithm}
\caption{\textsc{Cur-State-Dect-Ver-Obs}}
\label{alg:dect-tp}
\begin{algorithmic}[1]
\Statex \textbf{\!\!\!\!\!\!Inputs: }   $G$ and $\Sigma_F$
\Statex \textbf{\!\!\!\!\!\!Output: }  Current-State Detectable or Not
\State Build $V(G)=(X_V,  \Sigma_V  ,\delta_V  ,X_{0,V})$
\If {there exists a cycle $(x_0^1,x_0^2)(x_1^1,x_1^2)\dots (x_n^1,q_n^2)$ in $V(G)$ such that  $\exists i\!=\!0,\dots,n: x_i^1\!\not=\! x_i^2$}
\State \textbf{return} $G$ is not current-state detectable
\Else
 \State \textbf{return} $G$ is current-state detectable
\EndIf
\end{algorithmic}
\end{algorithm}

\begin{algorithm}
\caption{\textsc{Diag-Ver-TP}}
\label{alg:dia-tp}
\begin{algorithmic}[1]
\Statex \textbf{\!\!\!\!\!\!Inputs: }   $G$ and $\Sigma_F$
\Statex \textbf{\!\!\!\!\!\!Output: }  Diagnosable or Not
\State Build $V(G)=(X_V,  \Sigma_V  ,\delta_V  ,X_{0,V})$
\If {there exists a cycle $(x_0^1,x_0^2)(x_1^1,x_1^2)\dots (x_n^1,q_n^2)$ in $V(G)$ such that\\
 $\forall i=0,\dots,n: x_i^1\in X_N\wedge x_i^2\in X_F$}
\State \textbf{return} $G$ is not diagnosable
\Else
 \State \textbf{return} $G$ is  diagnosable
\EndIf
\end{algorithmic}
\end{algorithm}

\section{Related Problems and Further Readings}\label{sec:6}

\subsection{State Estimation under General Observation Models}
Throughout this article, we assume that the observation of the system is modeled as a natural projection.
The natural projection mapping is essentially static in the sense that an event is always either observable or unobservable.
One  related topic is the  \emph{sensor selection problem} for static observations \cite{haji1996minimizing,yoo2002np,jiang2003optimal,ru2010sensor,cabasino2013optimal},
i.e., we want to decide which events should be observable by placing with sensors such that a given observational property is fulfilled.
Another related topic in the static observation setting is the \emph{robust state estimation} problem when the observation is unreliable.
This problem has been investigated in the context of detectability analysis \cite{yin2017initial},
diagnosability anlaysis \cite{thorsley2008diagnosability,takai2012verification,carvalho2012robust,carvalho2013robust}
and supervisory control \cite{rohloff2005sensor,sanchez2006safe,lin2014control,alves2014robust,ushio2016nonblocking,yin2017supervisor} for partially-observed DES.

In many situations, due to information communications and acquisitions, the observation mapping may be \emph{dynamic},
i.e., whether or not an event is observable depends the trajectory of the system.
One example is the dynamic sensor activation problem, where we can decide to turn sensors on/off dynamically online based on the observation history.
In \cite{thorsley2007active,cassez2008fault,wang2010optimal,dallal2014most}, the fault diagnosis problem is studied under the dynamic observation setting.
In \cite{huang2008decentralized,wang2010minimization}, observability is studied under the dynamic observation setting.
In \cite{shu2010detectability,shu2013online}, the authors investigated the verification and synthesis of detectability under dynamic observations.
Opacity under dynamic observations is also studied in \cite{cassez2012synthesis,zhang2015max}. 
A general approach for dynamic sensor activation for property enforcement is proposed by \cite{yin2019general}. 
In \cite{wang2011codiagnosability,yin2015codiagnosability}, it has been shown that (co)diagnosability and (co)observability can be mapped from one to the other in the general dynamic observation setting. The reader is referred to the recent survey \cite{sears2016minimal} for more references on  state estimation problem under dynamic observations.

\subsection{State Estimation in Coordinated, Distributed and Modular Systems}
In the setting of this article, we only consider the scenario where the system is monitored by a single observer,  
which is referred to as the \emph{centralized state estimation}.
In many large-scale systems, sensors can be physically distributed and the system can be monitored by multiple local observers that have incomparable information.
Each local observer can perform local state estimation and send it to a coordinator.
Then the coordinator will fuses all local information according to some pre-specified protocol in order to obtain a global estimation decision.
This is referred to as the \emph{decentralized estimation and decision making problem under coordinated architecture}.

In the context of DES, the decentralized estimation and decision making problem was first studied by \cite{cieslak1988,rudie1992think,rudie1995computational} in the context of decentralized supervisory control, where the notion of coobservability is proposed.
In \cite{debouk2000coordinated,wang2007diagnosis}, the problem of decentralized fault diagnosis problem was studied and a corresponding property called codiagnosability was proposed; the verification codiagnosability  has also been studied in the literature \cite{moreira2011polynomial,qiu2006decentralized}.
The decentralized fault prognosis has also been studied in the literature; see, e.g., \cite{kumar2010decentralized,khoumsi2012conjunctive,yin2016decentralized2}.
Note that, in   decentralized decision making problems, one important issue is the underlying coordinated architecture.
For example, \cite{rudie1992think} and \cite{qiu2006decentralized} only consider  simple binary architectures  for control and diagnosis, respectively; more complicated architectures  can be found in \cite{yoo2002general,yoo2004decentralized,kumar2009inference,kumar2007inference,takai2011inference,chakib2011multi,ricker2007knowledge,yin2016decentralized,khoumsi2018decentralized}.

When each local observer is allow to communicate and exchange information with each other, the state estimation problem is referred to as the \emph{distributed} estimation problem.
Works on distributed state estimation and property verification can be founded in \cite{benveniste2003diagnosis,su2005global,genc2007distributed,takai2012distributed}.
Finally, state estimation and verification of partially-observed \emph{modular} DES have also been considered in the literature \cite{rohloff2005pspace,contant2006diagnosability,saboori2010reduced,schmidt2013verification,yin2017verification,masopust2019complexity},
where a modular system is composed by a set of local modules in the form of $G=G_1\times\cdots\times G_n$.

\subsection{Estimation and Verification of Petri Nets and Stochastic DES}
In this article, we focus on DES modeled as finite state automata.
Petri nets, another important class of DES models, are widely used to model many classes of concurrent systems.
In particular, Petri nets provide a compact model without enumerating the entire state space and it is well-known that Petri net languages are more expressive than regular languages.
The problems of state estimation and property verification have also drawn many attentions in the context of Petri nets.
For example, state (marking) estimation algorithms for Petri nets  have been  proposed  in \cite{giua2007marking,li2013minimum,bonhomme2015marking,basile2015state}.
Decidability and verification procedures for
diagnosability \cite{berard2018complexity,ran2018codiagnosability,basile2012k,cabasino2012new,yin2017decidability},
detectability \cite{masopust2018deciding,giua2002observability},
prognosability \cite{yin2018verification} and
opacity \cite{tong2017verification,tong2017decidability}
are also studied in the literature for Petri nets.

Another important generalization of the finite state automata model is the stochastic DES (or labelled Markov chains).
Stochastic DES can not only characterize whether or not a system can reach a state, it can also capture the possibility of reaching a state.
Hence it provides a model for the quantitative analysis and verification of DES.
State estimation and  verification of many important properties  have also been extended to the stochastic DES setting;
this includes, e.g., diagnosability \cite{yin2019robust,liu2008decentralized,thorsley2005diagnosability}, detectability \cite{shu2008state,keroglou2017verification,yin2017initial,zhao2019detectability}, prognosability \cite{chen2015stochastic} and
opacity \cite{saboori2014current,berard2015probabilistic,yin2019infinite}.

\subsection{Control Synthesis of Partially-Observed DES}

So far, we have only discussed  property verification problems in partially-observed DES.
In many applications, when  the answer to the verification problem is negative,  it is important to \emph{synthesize} a supervisor or controller that provably enforces the property by restricting the system behavior but as permissive as possible.
This control synthesis problem has been studied in the literature in the framework of the \emph{supervisory control theory} initiated by Ramadge and Wonham \cite{ramadge1987supervisory}. 
The reader is referred to the textbooks \cite{Lbook,Cbook} for more details on supervisory control of DES.
Supervisory control under partial observation  was originally investigated by \cite{cieslak1988,lin1988observability}. 
Since then, many control synthesis algorithms for partial-observation supervisors have been proposed \cite{cho1989supremal,kumar1991controllability,takai2003effective,yoo2006solvability,cai2015relative}.
In particular, \cite{yin2016synthesis} solves the synthesis problem for maximally-permissive non-blocking supervisors in the partial observation setting.

In the context of enforcement of observational properties, in \cite{sampath1998active}, an approach  was proposed for designing a  supervisor that enforces diagnosability.
Control synthesis algorithms have also been proposed in the literature for enforcing opacity; see, e.g. \cite{dubreil2010supervisory,tong2018current,saboori2012opacity}.
In \cite{shu2013enforcing}, the authors studied the problem of synthesizing   supervisors that enforce  detectability.
A uniform approach for control synthesis for enforcing a wide class of properties was recently proposed by Yin and Lafortune in a series of papers \cite{yin2018synthesis,yin2017synthesis,yin2016synthesis,yin2016uniform}.

\bibliographystyle{plain}
\bibliography{des}

\begin{thebibliography}{100}

\bibitem{alves2014robust}
M.V.S. Alves, J.C. Basilio, A.E. Carrilho~da Cunha, L.K. Carvalho, and M.V.
  Moreira.
\newblock Robust supervisory control against intermittent loss of observations.
\newblock In {\em 12th International Workshop on Discrete Event Systems},
  volume~12, pages 294--299, 2014.

\bibitem{badouel2007concurrent}
E.~Badouel, M.~Bednarczyk, A.~Borzyszkowski, B.~Caillaud, and P.~Darondeau.
\newblock Concurrent secrets.
\newblock {\em Discrete Event Dynamic Systems: Theory \& Applications},
  17(4):425--446, 2007.

\bibitem{basile2015state}
F.~Basile, M.P. Cabasino, and C.~Seatzu.
\newblock State estimation and fault diagnosis of labeled time petri net
  systems with unobservable transitions.
\newblock {\em IEEE Transactions on Automatic Control}, 60(4):997--1009, 2015.

\bibitem{basile2012k}
F.~Basile, P.~Chiacchio, and G.~De~Tommasi.
\newblock On ${K}$-diagnosability of {P}etri nets via integer linear
  programming.
\newblock {\em Automatica}, 48(9):2047--2058, 2012.

\bibitem{hadj2005characterizing}
N.~Ben Hadj-Alouane, S.~Lafrance, F.~Lin, J.~Mullins, and M.~Yeddes.
\newblock Characterizing intransitive noninterference for 3-domain security
  policies with observability.
\newblock {\em IEEE Transactions on Automatic Control}, 50(6):920--925, 2005.

\bibitem{hadj2005verification}
N.~Ben Hadj-Alouane, S.~Lafrance, F.~Lin, J.~Mullins, and M.~Yeddes.
\newblock On the verification of intransitive noninterference in mulitlevel
  security.
\newblock {\em IEEE Transactions on Systems, Man, and Cybernetics, Part B
  (Cybernetics)}, 35(5):948--958, 2005.

\bibitem{benveniste2003diagnosis}
A.~Benveniste, E.~Fabre, S.~Haar, and C.~Jard.
\newblock Diagnosis of asynchronous discrete-event systems: a net unfolding
  approach.
\newblock {\em IEEE Transactions on Automatic Control}, 48(5):714--727, 2003.

\bibitem{berard2015probabilistic}
B.~B{\'e}rard, K.~Chatterjee, and N.~Sznajder.
\newblock Probabilistic opacity for markov decision processes.
\newblock {\em Information Processing Letters}, 115(1):52--59, 2015.

\bibitem{berard2018complexity}
B.~B{\'e}rard, S.~Haar, S.~Schmitz, and S.~Schwoon.
\newblock The complexity of diagnosability and opacity verification for {P}etri
  nets.
\newblock {\em Fundamenta Informaticae}, 161(4):317--349, 2018.

\bibitem{bonhomme2015marking}
P.~Bonhomme.
\newblock Marking estimation of p-time {P}etri nets with unobservable
  transitions.
\newblock {\em IEEE Transactions on Systems, Man, and Cybernetics: Systems},
  45(3):508--518, 2015.

\bibitem{bryans2008opacity}
J.W. Bryans, M.~Koutny, L.~Mazar{\'e}, and P.Y.A. Ryan.
\newblock Opacity generalised to transition systems.
\newblock {\em International Journal of Information Security}, 7(6):421--435,
  2008.

\bibitem{cabasino2012new}
M.P. Cabasino, A.~Giua, S.~Lafortune, and C.~Seatzu.
\newblock A new approach for diagnosability analysis of {P}etri nets using
  verifier nets.
\newblock {\em IEEE Transactions on Automatic Control}, 57(12):3104--3117,
  2012.

\bibitem{cabasino2013optimal}
M.P. Cabasino, S.~Lafortune, and C.~Seatzu.
\newblock Optimal sensor selection for ensuring diagnosability in labeled
  {P}etri nets.
\newblock {\em Automatica}, 49(8):2373--2383, 2013.

\bibitem{cai2015relative}
K.~Cai, R.~Zhang, and W.M. Wonham.
\newblock Relative observability of discrete-event systems and its supremal
  sublanguages.
\newblock {\em IEEE Transactions on Automatic Control}, 60(3):659--670, 2015.

\bibitem{carvalho2013robust}
L.~K. Carvalho, M.V. Moreira, J.C. Basilio, and S.~Lafortune.
\newblock Robust diagnosis of discrete-event systems against permanent loss of
  observations.
\newblock {\em Automatica}, 49(1):223--231, 2013.

\bibitem{carvalho2012robust}
L.K. Carvalho, J.C. Basilio, and M.V. Moreira.
\newblock Robust diagnosis of discrete event systems against intermittent loss
  of observations.
\newblock {\em Automatica}, 48(9):2068--2078, 2012.

\bibitem{Lbook}
C.G. Cassandras and S.~Lafortune.
\newblock {\em Introduction to Discrete Event Systems}.
\newblock Springer, 2nd edition, 2008.

\bibitem{cassez2012synthesis}
F.~Cassez, J.~Dubreil, and H.~Marchand.
\newblock Synthesis of opaque systems with static and dynamic masks.
\newblock {\em Formal Methods in System Design}, 40(1):88--115, 2012.

\bibitem{cassez2008fault}
F.~Cassez and S.~Tripakis.
\newblock Fault diagnosis with static and dynamic observers.
\newblock {\em Fundamental Informaticae}, 88(4):497--540, 2008.

\bibitem{chakib2011multi}
H.~Chakib and A.~Khoumsi.
\newblock Multi-decision supervisory control: Parallel decentralized
  architectures cooperating for controlling discrete event systems.
\newblock {\em IEEE Transactions on Automatic Control}, 56(11):2608--2622,
  2011.

\bibitem{chen2015stochastic}
J.~Chen and R.~Kumar.
\newblock Stochastic failure prognosability of discrete event systems.
\newblock {\em IEEE Transactions on Automatic Control}, 60(6):1570--1581, 2015.

\bibitem{cho1989supremal}
H.~Cho and S.I. Marcus.
\newblock Supremal and maximal sublanguages arising in supervisor synthesis
  problems with partial observations.
\newblock {\em Mathematical Systems Theory}, 22(1):177--211, 1989.

\bibitem{cieslak1988}
R.~Cieslak, C.~Desclaux, A.S. Fawaz, and P.~Varaiya.
\newblock Supervisory control of discrete-event processes with partial
  observations.
\newblock {\em IEEE Transactions on Automatic Control}, 33(3):249--260, 1988.

\bibitem{contant2004diagnosis}
O.~Contant, S.~Lafortune, and D.~Teneketzis.
\newblock Diagnosis of intermittent faults.
\newblock {\em Discrete Event Dynamic Systems: Theory \& Applications},
  14(2):171--202, 2004.

\bibitem{contant2006diagnosability}
O.~Contant, S.~Lafortune, and D.~Teneketzis.
\newblock Diagnosability of discrete event systems with modular structure.
\newblock {\em Discrete Event Dynamic Systems: Theory \& Applications},
  16(1):9--37, 2006.

\bibitem{dallal2014most}
E.~Dallal and S.~Lafortune.
\newblock On most permissive observers in dynamic sensor activation problems.
\newblock {\em IEEE Transactions on Automatic Control}, 59(4):966--981, 2014.

\bibitem{debouk2000coordinated}
R.~Debouk, S.~Lafortune, and D.~Teneketzis.
\newblock Coordinated decentralized protocols for failure diagnosis of discrete
  event systems.
\newblock {\em Discrete Event Dynamic Systems: Theory \& Applications},
  10(1-2):33--86, 2000.

\bibitem{dubreil2010supervisory}
J.~Dubreil, P.~Darondeau, and H.~Marchand.
\newblock Supervisory control for opacity.
\newblock {\em IEEE Transactions on Automatic Control}, 55(5):1089--1100, 2010.

\bibitem{fabre2018diagnosability}
E.~Fabre, L.~H{\'e}lou{\"e}t, E.~Lefaucheux, and H.~Marchand.
\newblock Diagnosability of repairable faults.
\newblock {\em Discrete Event Dynamic Systems: Theory \& Applications},
  28(2):183--213, 2018.

\bibitem{genc2007distributed}
S.~Genc and S.~Lafortune.
\newblock Distributed diagnosis of place-bordered {P}etri nets.
\newblock {\em IEEE Transactions on Automation Science and Engineering},
  4(2):206--219, 2007.

\bibitem{genc2009predictability}
S.~Genc and S.~Lafortune.
\newblock Predictability of event occurrences in partially-observed
  discrete-event systems.
\newblock {\em Automatica}, 45(2):301--311, 2009.

\bibitem{giua2002observability}
A.~Giua and C.~Seatzu.
\newblock Observability of place/transition nets.
\newblock {\em IEEE Transactions on Automatic Control}, 47(9):1424--1437, 2002.

\bibitem{giua2007marking}
A.~Giua, C.~Seatzu, and D.~Corona.
\newblock Marking estimation of {P}etri nets with silent transitions.
\newblock {\em IEEE Transactions on Automatic Control}, 52(9):1695--1699, 2007.

\bibitem{hadjicostis2016k}
C.N. Hadjicostis and C.~Seatzu.
\newblock K-detectability in discrete event systems.
\newblock In {\em 55th IEEE Conference on Decision and Control}, pages
  420--425, 2016.

\bibitem{haji1996minimizing}
A.~Haji-Valizadeh and K.~Loparo.
\newblock Minimizing the cardinality of an events set for supervisors of
  discrete-event dynamical systems.
\newblock {\em IEEE Transactions on Automatic Control}, 41(11):1579--1593,
  1996.

\bibitem{huang2008decentralized}
Y.~Huang, K.~Rudie, and F.~Lin.
\newblock Decentralized control of discrete-event systems when supervisors
  observe particular event occurrences.
\newblock {\em IEEE Transactions on Automatic Control}, 53(1):384--388, 2008.

\bibitem{jacob2016overview}
R.~Jacob, J.-J. Lesage, and J.-M. Faure.
\newblock Overview of discrete event systems opacity: Models, validation, and
  quantification.
\newblock {\em Annual Reviews in Control}, 41:135--146, 2016.

\bibitem{jeron2008predictability}
T.~J{\'e}ron, H.~Marchand, S.~Genc, and S.~Lafortune.
\newblock Predictability of sequence patterns in discrete event systems.
\newblock {\em IFAC Proceedings Volumes}, 41(2):537--543, 2008.

\bibitem{jiang2001polynomial}
S.~Jiang, Z.~Huang, V.~Chandra, and R.~Kumar.
\newblock A polynomial algorithm for testing diagnosability of discrete-event
  systems.
\newblock {\em IEEE Transactions on Automatic Control}, 46(8):1318--1321, 2001.

\bibitem{jiang2003optimal}
S.~Jiang, R.~Kumar, and H.E Garcia.
\newblock Optimal sensor selection for discrete-event systems with partial
  observation.

\bibitem{jiang2003diagnosis}
S.~Jiang, R.~Kumar, and H.E. Garcia.
\newblock Diagnosis of repeated/intermittent failures in discrete event
  systems.
\newblock {\em IEEE Transactions on Robotics and Automation}, 19(2):310--323,
  2003.

\bibitem{keroglou2017verification}
C.~Keroglou and C.N. Hadjicostis.
\newblock Verification of detectability in probabilistic finite automata.
\newblock {\em Automatica}, 86:192--198, 2017.

\bibitem{khoumsi2012conjunctive}
A.~Khoumsi and H.~Chakib.
\newblock Conjunctive and disjunctive architectures for decentralized prognosis
  of failures in discrete-event systems.
\newblock {\em IEEE Transactions on Automation Science and Engineering},
  9(2):412--417, 2012.

\bibitem{khoumsi2018decentralized}
A.~Khoumsi and H.~Chakib.
\newblock Decentralized supervisory control of discrete event systems: An
  arborescent architecture to realize inference-based control.
\newblock {\em IEEE Transactions on Automatic Control}, 63(12):4278--4285,
  2018.

\bibitem{kumar1991controllability}
R.~Kumar, V.~Garg, and S.I. Marcus.
\newblock On controllability and normality of discrete event dynamical systems.
\newblock {\em Systems \& Control Letters}, 17(3):157--168, 1991.

\bibitem{kumar2007inference}
R.~Kumar and S.~Takai.
\newblock Inference-based ambiguity management in decentralized
  decision-making: Decentralized control of discrete event systems.
\newblock {\em IEEE Transactions on Automatic Control}, 52(10):1783--1794,
  2007.

\bibitem{kumar2009inference}
R.~Kumar and S.~Takai.
\newblock Inference-based ambiguity management in decentralized
  decision-making: Decentralized diagnosis of discrete-event systems.
\newblock {\em IEEE Transactions on Automation Science and Engineering},
  6(3):479--491, 2009.

\bibitem{kumar2010decentralized}
R.~Kumar and S.~Takai.
\newblock Decentralized prognosis of failures in discrete event systems.
\newblock {\em IEEE Transactions on Automatic Control}, 55(1):48--59, 2010.

\bibitem{li2013minimum}
L.~Li and C.N. Hadjicostis.
\newblock Minimum initial marking estimation in labeled {P}etri nets.
\newblock {\em IEEE Transactions on Automatic Control}, 58(1):198--203, 2013.

\bibitem{lin1994diagnosability}
F.~Lin.
\newblock Diagnosability of discrete event systems and its applications.
\newblock {\em Discrete Event Dynamic Systems: Theory \& Applications},
  4(2):197--212, 1994.

\bibitem{lin2011opacity}
F.~Lin.
\newblock Opacity of discrete event systems and its applications.
\newblock {\em Automatica}, 47(3):496--503, 2011.

\bibitem{lin2014control}
F.~Lin.
\newblock Control of networked discrete event systems: dealing with
  communication delays and losses.
\newblock {\em SIAM Journal onn Control and Optimization}, 52(2):1276--1298,
  2014.

\bibitem{lin1988observability}
F.~Lin and W.M. Wonham.
\newblock On observability of discrete-event systems.
\newblock {\em Information Sciences}, 44(3):173--198, 1988.

\bibitem{liu2008decentralized}
F.~Liu, D.~Qiu, H.~Xing, and Z.~Fan.
\newblock Decentralized diagnosis of stochastic discrete event systems.
\newblock {\em IEEE Transactions on Automatic Control}, 53(2):535--546, 2008.

\bibitem{masopust2018complexity}
T.~Masopust.
\newblock Complexity of deciding detectability in discrete event systems.
\newblock {\em Automatica}, 93:257--261, 2018.

\bibitem{masopust2019complexity}
T.~Masopust and X.~Yin.
\newblock Complexity of detectability, opacity and a-diagnosability for modular
  discrete event systems.
\newblock {\em Automatica}, 101:290--295, 2019.

\bibitem{masopust2018deciding}
T.~Masopust and X.~Yin.
\newblock Deciding detectability for labeled {P}etri nets.
\newblock {\em Automatica}, 2019.

\bibitem{mazare2004using}
L.~Mazar{\'e}.
\newblock Using unification for opacity properties.
\newblock In {\em Proceedings of WITS}, volume~4, pages 165--176, 2004.

\bibitem{moreira2011polynomial}
M.~V. Moreira, T.~C. Jesus, and J.~C. Basilio.
\newblock Polynomial time verification of decentralized diagnosability of
  discrete event systems.
\newblock {\em IEEE Transactions on Automatic Control}, 56(7):1679--1684, 2011.

\bibitem{ozveren1990observability}
C.M. Ozveren and A.S. Willsky.
\newblock Observability of discrete event dynamic systems.
\newblock {\em IEEE Transactions on Automatic Control}, 35(7):797--806, 1990.

\bibitem{qiu2006decentralized}
W.~Qiu and R.~Kumar.
\newblock Decentralized failure diagnosis of discrete event systems.
\newblock {\em IEEE Transactions on Systems, Man and Cybernetics, Part A},
  36(2):384--395, 2006.

\bibitem{ramadge1987supervisory}
P.J. Ramadge and W.M. Wonham.
\newblock Supervisory control of a class of discrete event processes.
\newblock {\em SIAM Journal on Control and Optimization}, 25(1):206--230, 1987.

\bibitem{ran2018codiagnosability}
N.~Ran, H.~Su, A.~Giua, and C.~Seatzu.
\newblock Codiagnosability analysis of bounded {P}etri nets.
\newblock {\em IEEE Transactions on Automatic Control}, 63(4):1192--1199, 2018.

\bibitem{ricker2007knowledge}
S.L. Ricker and K.~Rudie.
\newblock Knowledge is a terrible thing to waste: Using inference in
  discrete-event control problems.
\newblock {\em IEEE Transactions on Automatic Control}, 52(3):428--441, 2007.

\bibitem{rohloff2005sensor}
K.~Rohloff.
\newblock Sensor failure tolerant supervisory control.
\newblock In {\em 44th IEEE Conf.\ Decision and Control}, pages 3493--3498,
  2005.

\bibitem{rohloff2005pspace}
K.~Rohloff and S.~Lafortune.
\newblock {PSPACE}-completeness of modular supervisory control problems.
\newblock {\em Discrete Event Dynamic Systems: Theory \& Applications},
  15(2):145--167, 2005.

\bibitem{ru2010sensor}
Y.~Ru and C.N. Hadjicostis.
\newblock Sensor selection for structural observability in discrete event
  systems modeled by {P}etri nets.
\newblock {\em IEEE Transactions on Automatic Control}, 55(8):1751--1764, 2010.

\bibitem{rudie1995computational}
K.~Rudie and J.~C. Willems.
\newblock The computational complexity of decentralized discrete-event control
  problems.
\newblock {\em IEEE Transactions on Automatic Control}, 40(7):1313--1319, 1995.

\bibitem{rudie1992think}
K.~Rudie and W.~M. Wonham.
\newblock Think globally, act locally: Decentralized supervisory control.
\newblock {\em IEEE Transactions on Automatic Control}, 37(11):1692--1708,
  1992.

\bibitem{saboori2012opacity}
A.~Saboori and C.~N. Hadjicostis.
\newblock Opacity-enforcing supervisory strategies via state estimator
  constructions.
\newblock {\em IEEE Transactions on Automatic Control}, 57(5):1155--1165, 2012.

\bibitem{saboori2007notions}
A.~Saboori and C.N. Hadjicostis.
\newblock Notions of security and opacity in discrete event systems.
\newblock In {\em 46th IEEE Conference on Decision and Control}, pages
  5056--5061. IEEE, 2007.

\bibitem{saboori2010reduced}
A.~Saboori and C.N. Hadjicostis.
\newblock Reduced-complexity verification for initial-state opacity in modular
  discrete event systems.
\newblock {\em IFAC Proceedings Volumes}, 43(12):78--83, 2010.

\bibitem{saboori2011verification}
A.~Saboori and C.N. Hadjicostis.
\newblock Verification of ${K}$-step opacity and analysis of its complexity.
\newblock {\em IEEE Transactions on Automation Science and Engineering},
  8(3):549--559, 2011.

\bibitem{saboori2012verification}
A.~Saboori and C.N. Hadjicostis.
\newblock Verification of infinite-step opacity and complexity considerations.
\newblock {\em IEEE Transactions on Automatic Control}, 57(5):1265--1269, 2012.

\bibitem{saboori2013verification}
A.~Saboori and C.N. Hadjicostis.
\newblock Verification of initial-state opacity in security applications of
  discrete event systems.
\newblock {\em Information Sciences}, 246:115--132, 2013.

\bibitem{saboori2014current}
A.~Saboori and C.N. Hadjicostis.
\newblock Current-state opacity formulations in probabilistic finite automata.
\newblock {\em IEEE Transactions on Automatic Control}, 59(1):120--133, 2014.

\bibitem{sampath1998active}
M.~Sampath, S.~Lafortune, and D.~Teneketzis.
\newblock Active diagnosis of discrete-event systems.
\newblock {\em IEEE Transactions on Automatic Control}, 43(7):908--929, 1998.

\bibitem{sampath1995diagnosability}
M.~Sampath, R.~Sengupta, S.~Lafortune, K.~Sinnamohideen, and D.~Teneketzis.
\newblock Diagnosability of discrete-event systems.
\newblock {\em IEEE Transactions on Automatic Control}, 40(9):1555--1575, 1995.

\bibitem{sanchez2006safe}
A.M. S{\'a}nchez and F.J. Montoya.
\newblock Safe supervisory control under observability failure.
\newblock {\em Discrete Event Dynamic Systems: Theory \& Applications},
  16(4):493--525, 2006.

\bibitem{sasi2018detectability}
Y.~Sasi and F.~Lin.
\newblock Detectability of networked discrete event systems.
\newblock {\em Discrete Event Dynamic Systems: Theory \& Applications},
  28(3):449--470, 2018.

\bibitem{schmidt2013verification}
K.W. Schmidt.
\newblock Verification of modular diagnosability with local specifications for
  discrete-event systems.
\newblock {\em IEEE Transactions on Systems, Man, and Cybernetics: Systems},
  43(5):1130--1140, 2013.

\bibitem{sears2014computing}
D.~Sears and K.~Rudie.
\newblock On computing indistinguishable states of nondeterministic finite
  automata with partially observable transitions.
\newblock In {\em 53rd IEEE Conference on Decision and Control}, pages
  6731--6736, 2014.

\bibitem{sears2016minimal}
D.~Sears and K.~Rudie.
\newblock Minimal sensor activation and minimal communication in discrete-event
  systems.
\newblock {\em Discrete Event Dynamic Systems: Theory \& Applications},
  26(2):295--349, 2016.

\bibitem{shu2013online}
S.~Shu, Z.~Huang, and F.~Lin.
\newblock Online sensor activation for detectability of discrete event systems.
\newblock {\em IEEE Trans. Automation Science and Engineering}, 10(2):457--461,
  2013.

\bibitem{shu2010detectability}
S.~Shu and F.~Lin.
\newblock Detectability of discrete event systems with dynamic event
  observation.
\newblock {\em Systems \& Control Letters}, 59(1):9--17, 2010.

\bibitem{shu2011generalized}
S.~Shu and F.~Lin.
\newblock Generalized detectability for discrete event systems.
\newblock {\em Systems \& control letters}, 60(5):310--317, 2011.

\bibitem{shu2013delayed}
S.~Shu and F.~Lin.
\newblock Delayed detectability of discrete event systems.
\newblock {\em IEEE Transactions on Automatic Control}, 58(4):862--875, 2013.

\bibitem{shu2013enforcing}
S.~Shu and F.~Lin.
\newblock Enforcing detectability in controlled discrete event systems.
\newblock {\em IEEE Transactions on Automatic Control}, 58(8):2125--2130, 2013.

\bibitem{shu2013detectability}
S.~Shu and F.~Lin.
\newblock I-detectability of discrete-event systems.
\newblock {\em IEEE Transactions on Automation Science and Engineering},
  10(1):187--196, 2013.

\bibitem{shu2007detectability}
S.~Shu, F.~Lin, and H.~Ying.
\newblock Detectability of discrete event systems.
\newblock {\em IEEE Transactions on Automatic Control}, 52(12):2356--2359,
  2007.

\bibitem{shu2008state}
S.~Shu, F.~Lin, H.~Ying, and X.~Chen.
\newblock State estimation and detectability of probabilistic discrete event
  systems.
\newblock {\em Automatica}, 44(12):3054--3060, 2008.

\bibitem{su2005global}
R.~Su and W.M. Wonham.
\newblock Global and local consistencies in distributed fault diagnosis for
  discrete-event systems.
\newblock {\em IEEE Transactions on Automatic Control}, 50(12):1923--1935,
  2005.

\bibitem{sweeney2002k}
L.~Sweeney.
\newblock $k$-anonymity: A model for protecting privacy.
\newblock {\em International Journal of Uncertainty, Fuzziness and
  Knowledge-Based Systems}, 10(05):557--570, 2002.

\bibitem{takai2011inference}
S.~Takai and R.~Kumar.
\newblock Inference-based decentralized prognosis in discrete event systems.
\newblock {\em IEEE Transactions on Automatic Control}, 56(1):165--171, 2011.

\bibitem{takai2012distributed}
S.~Takai and R.~Kumar.
\newblock Distributed failure prognosis of discrete event systems with
  bounded-delay communications.
\newblock {\em IEEE Transactions on Automatic Control}, 57(5):1259--1265, 2012.

\bibitem{takai2003effective}
S.~Takai and T.~Ushio.
\newblock Effective computation of an $l_m (g)$-closed, controllable, and
  observable sublanguage arising in supervisory control.
\newblock {\em Systems \& Control Letters}, 49(3):191--200, 2003.

\bibitem{takai2012verification}
S.~Takai and T.~Ushio.
\newblock Verification of codiagnosability for discrete event systems modeled
  by mealy automata with nondeterministic output functions.
\newblock {\em IEEE Transactions on Automatic Control}, 57(3):798--804, 2012.

\bibitem{thorsley2005diagnosability}
D.~Thorsley and D.~Teneketzis.
\newblock Diagnosability of stochastic discrete-event systems.
\newblock {\em IEEE Transactions on Automatic Control}, 50(4):476--492, 2005.

\bibitem{thorsley2007active}
D.~Thorsley and D.~Teneketzis.
\newblock Active acquisition of information for diagnosis and supervisory
  control of discrete event systems.
\newblock {\em Discrete Event Dynamic Systems: Theory \& Applications},
  17(4):531--583, 2007.

\bibitem{thorsley2008diagnosability}
D.~Thorsley, T.-S. Yoo, and H.E. Garcia.
\newblock Diagnosability of stochastic discrete-event systems under unreliable
  observations.
\newblock In {\em American Control Conference}, pages 1158--1165, 2008.

\bibitem{tong2017decidability}
Y.~Tong, Z.~Li, C.~Seatzu, and A.~Giua.
\newblock Decidability of opacity verification problems in labeled {P}etri net
  systems.
\newblock {\em Automatica}, 80:48--53, 2017.

\bibitem{tong2017verification}
Y.~Tong, Z.~Li, C.~Seatzu, and A.~Giua.
\newblock Verification of state-based opacity using {P}etri nets.
\newblock {\em IEEE Transactions on Automatic Control}, 62(6):2823--2837, 2017.

\bibitem{tong2018current}
Y.~Tong, Z.~Li, C.~Seatzu, and A.~Giua.
\newblock Current-state opacity enforcement in discrete event systems under
  incomparable observations.
\newblock {\em Discrete Event Dynamic Systems: Theory \& Applications},
  28(2):161--182, 2018.

\bibitem{tsitsiklis1989control}
J.N Tsitsiklis.
\newblock On the control of discrete-event dynamical systems.
\newblock {\em Mathematics of Control, Signals and Systems}, 2(2):95--107,
  1989.

\bibitem{ushio2016nonblocking}
T.~Ushio and S.~Takai.
\newblock Nonblocking supervisory control of discrete event systems modeled by
  mealy automata with nondeterministic output functions.
\newblock {\em IEEE Transactions on Automatic Control}, 61(3):799--804, 2016.

\bibitem{wang2011codiagnosability}
W.~Wang, A.~R. Girard, S.~Lafortune, and F.~Lin.
\newblock On codiagnosability and coobservability with dynamic observations.
\newblock {\em IEEE Transactions on Automatic Control}, 56(7):1551--1566, 2011.

\bibitem{wang2010optimal}
W.~Wang, S.~Lafortune, A.~R. Girard, and F.~Lin.
\newblock Optimal sensor activation for diagnosing discrete event systems.
\newblock {\em Automatica}, 46(7):1165--1175, 2010.

\bibitem{wang2007algorithm}
W.~Wang, S.~Lafortune, and F.~Lin.
\newblock An algorithm for calculating indistinguishable states and clusters in
  finite-state automata with partially observable transitions.
\newblock {\em Systems \& Control Letters}, 56(9-10):656--661, 2007.

\bibitem{wang2010minimization}
W.~Wang, S.~Lafortune, F.~Lin, and A.~R. Girard.
\newblock Minimization of dynamic sensor activation in discrete event systems
  for the purpose of control.
\newblock {\em IEEE Transactions on Automatic Control}, 55(11):2447--2461,
  2010.

\bibitem{wang2007diagnosis}
Y.~Wang, T.-S. Yoo, and S.~Lafortune.
\newblock Diagnosis of discrete event systems using decentralized
  architectures.
\newblock {\em Discrete Event Dynamic Systems: Theory \& Applications},
  17(2):233--263, 2007.

\bibitem{Cbook}
W.M. Wonham and K.~Cai.
\newblock {\em Supervisory Control of Discrete-Event Systems}.
\newblock Springer, 2019.

\bibitem{wu2013comparative}
Y.-C. Wu and S.~Lafortune.
\newblock Comparative analysis of related notions of opacity in centralized and
  coordinated architectures.
\newblock {\em Discrete Event Dynamic Systems: Theory \& Applications},
  23(3):307--339, 2013.

\bibitem{yin2017initial}
X.~Yin.
\newblock Initial-state detectability of stochastic discrete-event systems with
  probabilistic sensor failures.
\newblock {\em Automatica}, 80:127--134, 2017.

\bibitem{yin2017supervisor}
X.~Yin.
\newblock Supervisor synthesis for mealy automata with output functions: A
  model transformation approach.
\newblock {\em IEEE Transactions on Automatic Control}, 62(5):2576--2581, 2017.

\bibitem{yin2018verification}
X.~Yin.
\newblock Verification of prognosability for labeled {P}etri nets.
\newblock {\em IEEE Transactions on Automatic Control}, 63(6):1828--1834, 2018.

\bibitem{yin2019robust}
X.~Yin, J.~Chen, Z.~Li, and S.~Li.
\newblock Robust fault diagnosis of stochastic discrete event systems.
\newblock {\em IEEE Transactions on Automatic Control}, 2019.

\bibitem{yin2015codiagnosability}
X.~Yin and S.~Lafortune.
\newblock Codiagnosability and coobservability under dynamic observations:
  Transformation and verification.
\newblock {\em Automatica}, 61:241--252, 2015.

\bibitem{yin2016decentralized2}
X.~Yin and S.~Lafortune.
\newblock Decentralized supervisory control with intersection-based
  architecture.
\newblock {\em IEEE Transactions on Automatic Control}, 61(11):3644--3650,
  2016.

\bibitem{yin2016synthesis}
X.~Yin and S.~Lafortune.
\newblock Synthesis of maximally permissive supervisors for partially-observed
  discrete-event systems.
\newblock {\em IEEE Transactions on Automatic Control}, 61(5):1239--1254, 2016.

\bibitem{yin2016uniform}
X.~Yin and S.~Lafortune.
\newblock A uniform approach for synthesizing property-enforcing supervisors
  for partially-observed discrete-event systems.
\newblock {\em IEEE Transactions on Automatic Control}, 61(8):2140--2154, 2016.

\bibitem{yin2017new}
X.~Yin and S.~Lafortune.
\newblock A new approach for the verification of infinite-step and k-step
  opacity using two-way observers.
\newblock {\em Automatica}, 80:162--171, 2017.

\bibitem{yin2017decidability}
X.~Yin and S.~Lafortune.
\newblock On the decidability and complexity of diagnosability for labeled
  {P}etri nets.
\newblock {\em IEEE Transactions on Automatic Control}, 62(11):5931--5938,
  2017.

\bibitem{yin2017synthesis}
X.~Yin and S.~Lafortune.
\newblock Synthesis of maximally-permissive supervisors for the range control
  problem.
\newblock {\em IEEE Transactions on Automatic Control}, 62(8):3914--3929, 2017.

\bibitem{yin2017verification}
X.~Yin and S.~Lafortune.
\newblock Verification complexity of a class of observational properties for
  modular discrete events systems.
\newblock {\em Automatica}, 83:199--205, 2017.

\bibitem{yin2018minimization}
X.~Yin and S.~Lafortune.
\newblock Minimization of sensor activation in decentralized discrete-event
  systems.
\newblock {\em IEEE Transactions on Automatic Control}, 63(11):3705--3718,
  2018.

\bibitem{yin2018synthesis}
X.~Yin and S.~Lafortune.
\newblock Synthesis of maximally permissive nonblocking supervisors for the
  lower bound containment problem.
\newblock {\em IEEE Transactions on Automatic Control}, 63(12):4435--4441,
  2018.

\bibitem{yin2019general}
X.~Yin and S.~Lafortune.
\newblock A general approach for optimizing dynamic sensor activations for
  discrete event systems.
\newblock {\em Automatica}, 2019.

\bibitem{yin2016decentralized}
X.~Yin and Z.~Li.
\newblock Decentralized fault prognosis of discrete event systems with
  guaranteed performance bound.
\newblock {\em Automatica}, 69:375--379, 2016.

\bibitem{yin2019infinite}
X.~Yin, Z.~Li, W.~Wang, and S.~Li.
\newblock Infinite-step opacity and ${K}$-step opacity of stochastic
  discrete-event systems.
\newblock {\em Automatica}, 99:266--274, 2019.

\bibitem{yoo2004decentralized}
S.~Yoo, T.-S.and~Lafortune.
\newblock Decentralized supervisory control with conditional decisions:
  Supervisor existence.
\newblock {\em IEEE Transactions on Automatic Control}, 49(11):1886--1904,
  2004.

\bibitem{yoo2008diagnosis}
T.-S. Yoo and H.E. Garcia.
\newblock Diagnosis of behaviors of interest in partially-observed
  discrete-event systems.
\newblock {\em Systems \& Control Letters}, 57(12):1023--1029, 2008.

\bibitem{yoo2002general}
T.-S. Yoo and S.~Lafortune.
\newblock A general architecture for decentralized supervisory control of
  discrete-event systems.
\newblock {\em Discrete Event Dynamic Systems: Theory \& Applications},
  12(3):335--377, 2002.

\bibitem{yoo2002np}
T.-S. Yoo and S.~Lafortune.
\newblock Np-completeness of sensor selection problems arising in partially
  observed discrete-event systems.
\newblock {\em IEEE Transactions on Automatic Control}, 47(9):1495--1499, 2002.

\bibitem{yoo2002polynomial}
T.-S. Yoo and S.~Lafortune.
\newblock Polynomial-time verification of diagnosability of partially observed
  discrete-event systems.
\newblock {\em IEEE Transactions on Automatic Control}, 47(9):1491--1495, 2002.

\bibitem{yoo2006solvability}
T.-S. Yoo and S.~Lafortune.
\newblock Solvability of centralized supervisory control under partial
  observation.
\newblock {\em Discrete Event Dynamic Systems: Theory \& Applications},
  16(4):527--553, 2006.

\bibitem{zad2003fault}
S.H. Zad, R.H. Kwong, and W.M. Wonham.
\newblock Fault diagnosis in discrete-event systems: Framework and model
  reduction.
\newblock {\em IEEE Transactions on Automatic Control}, 48(7):1199--1212, 2003.

\bibitem{zaytoon2013overview}
J.~Zaytoon and S.~Lafortune.
\newblock Overview of fault diagnosis methods for discrete event systems.
\newblock {\em Annual Reviews in Control}, 37(2):308--320, 2013.

\bibitem{zhang2015max}
B.~Zhang, S.~Shu, and F.~Lin.
\newblock Maximum information release while ensuring opacity in discrete event
  systems.
\newblock {\em IEEE Transactions on Automation Science and Engineering},
  12(4):1067--1079, 2015.

\bibitem{zhang2017problem}
K.~Zhang.
\newblock The problem of determining the weak (periodic) detectability of
  discrete event systems is pspace-complete.
\newblock {\em Automatica}, 81:217--220, 2017.

\bibitem{zhao2019detectability}
P.~Zhao, S.~Shu, F.~Lin, and B.~Zhang.
\newblock Detectability measure for state estimation of discrete event systems.
\newblock {\em IEEE Transactions on Automatic Control}, 64(1):433--439, 2019.

\end{thebibliography}

\end{document}